\documentclass[english]{article}
\usepackage[T1]{fontenc}
\usepackage[latin9]{inputenc}
\usepackage{geometry}
\usepackage{float}
\usepackage{amsmath}
\usepackage{graphicx}
\usepackage{setspace}
\usepackage{esint}
\makeatletter
\newcommand\der{\mathrm{d}}    
\newcommand\CJ{\mathrm{CJ}}  

\makeatother

\usepackage{babel}
\begin{document}

\title{Set-valued solutions for non-ideal detonation}

\author{R. Semenko$^{*}$, L. Faria$^{*}$, A. Kasimov%
\thanks{Applied Mathematics and Computational Sciences, King Abdullah University
of Science and Technology, Thuwal, Saudi Arabia%
} %
\thanks{Corresponding author, aslan.kasimov@kaust.edu.sa%
}, and B. Ermolaev%
\thanks{Semenov Insitute of Chemical Physics, Moscow, Russia%
}}
\maketitle
\begin{abstract}
The existence and structure of steady gaseous detonation propagating
in a packed bed of solid inert particles are analyzed in the one-dimensional
approximation by taking into consideration frictional and heat losses
between the gas and the particles. A new formulation of the governing
equations is introduced that eliminates the well-known difficulties
with numerical integration across the sonic singularity in the reactive
Euler equations. The new algorithm allows us to determine that the
detonation solutions as the loss factors are varied have a set-valued
nature at low detonation velocities when the sonic constraint disappears
from the solutions. These set-valued solutions correspond to a continuous
spectrum of the eigenvalue problem that determines the velocity of
the detonation.
\end{abstract}

\section{Introduction}

Detonation is a shock wave in a reactive medium that is sustained
by the energy released in the chemical reactions initiated by the
passage of the wave. Gaseous detonation propagating in a tube, in
the interstitial space in a porous medium, or other obstructed environments
is subject to resistance due to heat transfer, boundary layers and
shock reflections that arise at the interfaces between the gas and
the obstacles or the walls of the tube. Such resistance leads to momentum
and energy losses in the gas that can significantly affect the propagation
of the detonation. In the presence of such losses, the detonation
is termed ``non-ideal'', and its propagation velocity is lower than
in the case (termed ``ideal'') when there are no such losses. This
drop in the detonation velocity is called the ``velocity deficit''.
Quantifying the velocity deficit in terms of the losses is one of
the major problems in detonation theory and its study goes back to
the original paper by Zel'dovich \cite{Zeldovich1940} (see also the
classic book by Zel'dovich and Kompaneets \cite{zeldovich1960theory}).
An important feature observed in experiments on non-ideal detonation
waves is their ability to propagate with velocities ranging from near
sonic speed in the ambient gas, $c_{a}$, to the ideal Chapman-Jouguet
(CJ) value, $D_{\CJ}$ \cite{lee1985turbulent,lyamin1986combustion,pinaev1989fundamental,lyamin1991propagation}.
The physical mechanisms responsible for this self-sustained propagation
of detonation depend on the extent and the nature of the losses. A
particularly interesting aspect is the possibility that the frictional
heating of the gas dominates over its heating by chemical reactions
at low detonation velocities. We also mention a peculiar theoretical
observation, first discussed in \cite{Zeldovich1940}, that in the
presence of losses, the flow in the detonation reaction zone can reverse
its direction; that is, the velocity can become negative in the laboratory
frame of reference, a situation typical for flames, but unusual for
detonations. No experimental evidence for this prediction appears
to exist. 

Since the early works mentioned above, this problem has been revisited
many times by researchers seeking to understand the precise mechanisms
of detonation propagation when the effects of losses are substantial.
A particularly important issue, from a practical point of view, is
the existence of detonation limits. The limit of detonation propagation
is defined as a threshold in terms of some control quantity (e.g.,
a tube diameter) below which the detonation cannot propagate. The
problem of defining such limits is nontrivial and is complicated by
the fact that detonation can and most often does propagate unsteadily.
Although this fact makes the definitions of these limits based on
the non-existence of stationary solutions somewhat limited in application,
understanding the steady-state solutions is a major first step in
understanding the nature of detonation limits as well as in building
a theory of unsteady detonations.

Theoretical work on the role of losses in detonations was continued
in \cite{zel1987detonation,zel1988nonideal,brailovsky2000hydraulic-a,brailovsky2002effects}.
In \cite{zel1987detonation}, both convective and conductive heat
losses were considered. Convective losses depend on the relative velocity
between the gas and the particles/walls and they vanish as the relative
velocity vanishes. In contrast, conductive losses persist in the absence
of the relative velocity. In the relaxation zone far downstream of
the lead shock, where the flow of the reaction products has sufficiently
decelerated, the authors of \cite{zel1987detonation} assumed that
the heat losses corresponded to an approximately constant Nusselt
number and thus were of the conductive type. Subsequently, they reasoned
that the gas in the products cooled down due to the conductive heat
transfer and reached a state of rest far downstream but with the same
pressure and density as the upstream state with the fresh mixture. 

In contrast, the model we present here involves only convective heat
losses, as a consequence of which the heat transfer from the gas to
the particles vanishes when the gas stops. This is a reasonable assumption
for detonations because the conductive losses occur on time scales
that are much longer than the time scale of the detonation propagation.
The temperature relaxation in the products can therefore not be expected
to affect the detonation dynamics over the time scales of interest.
We should note that our governing equations take the same general
form as those in \cite{Zeldovich1940,zeldovich1960theory,zel1987detonation}
and include the momentum loss due to friction, heat loss, which is
proportional to the temperature difference between the gas and the
solid particles, and the effect of heating of the gas due to the friction. 

Convective heat transfer was also considered in \cite{brailovsky2002effects}.
The heat loss term in the energy equation in \cite{brailovsky2002effects}
is slightly different from that of \cite{Zeldovich1940,zeldovich1960theory,zel1987detonation}
and hence also different from our model. In addition and more importantly,
the far-field conditions in \cite{brailovsky2002effects} include
a prescribed temperature value. This prescribed condition on the product
temperature follows directly from the energy equation when the heat
losses are neglected \cite{brailovsky2000hydraulic-a}. However, in
the presence of heat losses, imposing a prescribed temperature of
the products is difficult to justify. The only condition that we impose
here is that of the vanishing velocity in the products. The product
temperature is therefore an outcome of the solution of the steady-state
equations, not something prescribed. As a result, this formulation
has an interesting consequence: a new class of solutions with a continuous
spectrum of detonation velocities arises.

To highlight the most important general features of detonation in
systems with losses that have been found previously, we mention that,
in qualitative agreement with prior experimental research \cite{lee1985turbulent,lyamin1986combustion,pinaev1989fundamental,lyamin1991propagation},
different regimes of detonation propagation have been theoretically
identified in \cite{brailovsky2000hydraulic-a,brailovsky2002effects}: 

1) When the detonation speed is below the ideal CJ value but not substantially,
the wave has a structure that is very similar to that of the classical
ZND theory (Zel'dovich \cite{Zeldovich1940}, von Neumann \cite{vonNeumann1942}
and Döring \cite{Doering1943}; see also \cite{fickett2011detonation})
of a self-sustained detonation with its inherent transonic character
in the post-shock flow and with a sonic point located at some distance
from the shock.%
\footnote{Such detonation has been called ``quasi-detonation'' in the past
\cite{lee1985turbulent,brailovsky2000hydraulic-a,brailovsky2002effects}.
However, this term is somewhat misleading, as it implies that non-ideal
detonation is essentially distinct from classical ideal detonation.
In fact, non-ideal detonation is still a detonation wave, understood
as a self-sustained process of shock-wave reaction-zone coupling.
The presence of losses does not change this essential nature of the
wave as long as there is a sonic confinement in the flow.%
} 

2) When the detonation speed drops significantly, often to around
$0.6$ of the ideal value, the sonic locus moves to infinity and,
with a further decrease of the detonation velocity, disappears from
the flow altogether. This structure, in which the post-shock flow
is entirely subsonic in the shock-fixed frame, resembles piston-supported
overdriven detonations. However, the latter gain energy from the piston
and therefore propagate faster than CJ detonations. In contrast, non-ideal
detonation, with a subsonic post-shock flow, has no support and propagates
at velocities that are substantially lower than the ideal CJ velocity.
Because its structure is missing a sonic point, it is not a self-sustained
wave%
\footnote{This regime was termed a ``chocking regime'' in \cite{lee1985turbulent,brailovsky2000hydraulic-a,brailovsky2002effects}.
However, this is an unfortunate term because the flow in this case
does not contain a sonic point. Recall, that a choking regime in fluid
dynamics is associated with transonic flows.%
}.

In experiments, the velocity deficits are obtained by, for example,
decreasing the initial pressure, $p_{a}$, in the explosive mixture
\cite{lee1985turbulent,lyamin1986combustion,pinaev1989fundamental,lyamin1991propagation}.
In many cases, a decrease in $p_{a}$ to some critical value results
in an abrupt decrease in the speed to subsonic velocities, which is
more closely associated with turbulent deflagration waves than with
detonations. In theoretical modeling, one can, in principle, look
at the solutions that have even smaller velocity than the ambient
sound speed, as was done in \cite{brailovsky2000hydraulic-a,brailovsky2002effects}
and in subsequent works by the same authors and their collaborators.
However, in such cases, the structure does not contain a shock and
the wave is subsonic relative to the state upstream. We limit the
discussion in this work to supersonic waves containing a shock as
an inherent element of the structure.

While the nature of self-sustained non-ideal detonation is relatively
clear, that of the regime with subsonic post-shock flow has led to
some conflicting conclusions in the literature. In particular, the
existence of steady solutions of the governing equations in this regime
was questioned in \cite{dionne2000transient}. Indeed, because the
sonic point disappears from the flow, the classical self-sustained
structure cannot be a solution. It therefore appears reasonable to
conclude that no steady-state solutions exist in this case. In \cite{brailovsky2000hydraulic-a,brailovsky2002effects},
however, the authors have been able to construct the steady $D$-$c_{f}$
relations in the entire range of velocities from the ideal $D=D_{\CJ}$
value to the sonic velocity, $D=c_{a}$ (and even further below),
where $c_{a}$ is the speed of sound in the upstream mixture and $c_{f}$
is the dimensionless drag coefficient. Instead of the sonic conditions,
they applied $u=0$ at infinity and solved the appropriate boundary
value problem. In this connection, a point that requires certain care
in interpretation is the question of the existence of a steady-state
solution versus its stability. Numerical simulations in \cite{dionne2000transient,Ermolaev-2011}
indicate that detonation with frictional momentum losses is unstable;
in fact, it is more unstable in the presence of losses than in their
absence. Because of the instability, the steady-state solutions cannot
be reached as long-time asymptotic solutions of an initial value problem
of the underlying reactive Euler equations. This, however, does not
imply that the steady-state solutions of the equations do not exist.
In order to establish the existence of the steady-state solutions,
we must be able to solve the appropriate boundary value problem for
the time-independent governing equations. Importantly, the proper
statement of such a boundary value problem does not necessarily involve
sonic conditions. We discuss this point in detail in the subsequent
sections of this paper. 

In this work, we show that the steady-state solutions of the governing
reactive Euler equations with heat and momentum losses have a peculiar
nature in that the $D$-$c_{f}$ dependence is no longer a curve,
but rather a set-valued function. When the sonic locus disappears
from the flow, at a given loss coefficient, there exists a continuous
range of detonation velocities, all corresponding to steady-state
solutions of the governing equations. Effectively, $D$, as the eigenvalue
of the nonlinear eigenvalue problem, is found to have both discrete
and continuous spectra. This is in contrast with prior results, in
which either the steady-state solution appears not to exist \cite{dionne2000transient}
or that there is a Z-shaped curve in the $D$-$c_{f}$ plane, giving
a finite and discrete set of solutions for a given $c_{f}$ \cite{higgins2012steady}. 

Another contribution of our work is a new formulation of the governing
equations that completely avoids the difficulty of integrating through
the sonic locus. The latter is a well-recognized problem in detonation
theory \cite{zel1987detonation,higgins2012steady,bdzil2012theory}.
We have discovered a new set of dependent variables in which the governing
equations are no longer singular at the sonic point. This finding
results in a numerically well-conditioned integration of the detonation
structure in the entire flow region. Generalization of the algorithm
to incorporate other loss factors and multiple reactions is possible
and is introduced in \cite{faria2013}. 

A possibility of flow reversal in the reaction zone \cite{Zeldovich1940}
was also briefly discussed in \cite{zel1987detonation}, where the
authors continued the analysis of the problem and found that, with
the far-field conditions they used, the convective heat transfer between
the gas and the tube walls (or solid particles) was insufficient to
cause the flow reversal. However, with the inclusion of conductive
heat losses, flow reversal was indeed found. The authors of \cite{brailovsky2002effects}
did not discuss this possibility even though convective heat transfer
was also an essential element of their analysis. Our present finding
is that, with convective heat transfer, flow reversal necessarily
occurs in the regime where the flow behind the shock is entirely subsonic.
In contrast, no flow reversal is possible with only friction losses.
We emphasize, however, that our far-field conditions are different
from those of \cite{zel1987detonation} and \cite{brailovsky2002effects}.

The remainder of this paper is organized as follows. In Section \ref{sec:Governing-equations},
we introduce reactive Euler equations that are subject to the momentum
and heat exchange terms. The Rankine-Hugoniot conditions are also
introduced in this section. In Section \ref{sec:Steady-state-solution},
we discuss the steady-state solutions and introduce a change of dependent
variables that removes the sonic singularities from the equations.
In Section \ref{sec:Example-calculation}, we calculate various solutions
when both the friction and heat loss terms are present and show that
the detonation speed versus loss coefficient is a set-valued function.
Concluding remarks are offered in Section \ref{sec:Conclusions}.

\section{\label{sec:Governing-equations}Governing equations}

Consider a detonation in a perfect gas that fills the interstitial
space in a packed bed of solid particles of diameter $d$. The particles
are assumed to be immobile and inert; their only role is to exchange
momentum and energy with the detonating gas. The chemical reaction
in the gas is described globally as $Reactant\to Product$. The progress
of this reaction is measured by variable, $\lambda$, that goes from
$0$ at the shock to $1$ in the fully burnt gas. The reaction rate
is assumed to be given in the Arrhenius form, 
\begin{equation}
\omega=k\left(1-\lambda\right)\exp\left(-\frac{E}{pv}\right),
\end{equation}
where $E$ is the activation energy, $p$ is the pressure, and $v$
is the specific volume. The perfect-gas equation of state is assumed
to be given by 
\begin{equation}
e_{i}=\frac{pv}{\gamma-1},
\end{equation}
where $\gamma$ is the constant ratio of specific heats.

With these modeling assumptions, the governing reactive Euler equations
further incorporating the heat and momentum exchange between the gas
and particles are as follows. The continuity equation is 

\begin{equation}
\rho_{t}+(\rho u)_{x}=0,\label{eq:continuity}
\end{equation}
where $\rho=1/v$ and $u$ are the gas density and velocity, respectively.
The gas momentum equation is 
\begin{equation}
u_{t}+uu_{x}=-\frac{1}{\rho}p_{x}-\frac{f}{\rho\phi},
\end{equation}
or in conservation form, 
\begin{equation}
\left(\rho u\right)_{t}+\left(p+\rho u^{2}\right)_{x}=-\frac{f}{\phi},\label{eq:momentum}
\end{equation}
where the drag force due to the solid particles is assumed to take
the form \cite{goldshtik1984transfer}, 
\begin{equation}
f=A_{s}\rho\left(b_{1}+\frac{b_{2}}{Re}\right)u|u|,
\end{equation}
with 
\begin{equation}
A_{s}=\frac{4\phi}{d_{p}},\quad d_{p}=\frac{2\phi d}{3(1-\phi)},\quad\mathrm{Re}=\frac{d_{p}|u|}{\nu},
\end{equation}
where $\phi$ is the porosity of the packed bed, i.e., the fraction
of space occupied by the gas, $d$ is the particle diameter, $\mathrm{Re}$
is the Reynolds number, $\nu$ is the gas kinematic viscosity, $b_{1}$
and $b_{2}$ are numerical constants. In the calculations below, we
take $\phi=0.4$, $b_{1}=0.75$, $b_{2}=0$ \cite{goldshtik1984transfer},
thereby assuming high Reynolds number friction losses.

The energy equation can be written as 
\begin{equation}
p_{t}+up_{x}+\gamma pu_{x}=\left(\gamma-1\right)Q\rho\omega+\left(\gamma-1\right)\left(\frac{uf-h}{\phi}\right).\label{eq:energy_p}
\end{equation}
It contains contributions due to: 1) the chemical energy release (the
first term, wherein $Q$ is the heat release), 2) the work done by
the friction forces (the second term, involving $uf$), and 3) the
heat transfer between the gas and particles (the last term, involving
$h$). The heat exchange rate, $h$, is given by Newton's law, 
\begin{equation}
h=A_{s}\alpha_{s}\left(T-T_{s}\right),
\end{equation}
in which the particle temperature is denoted by $T_{s}$ and the heat
conduction coefficient, $\alpha_{s}$, is calculated from \cite{goldshtik1984transfer}
\begin{equation}
\alpha_{s}=\lambda_{g}\frac{\mathrm{Nu}}{d_{p}},
\end{equation}
where $\mathrm{Nu}=a_{1}+a_{2}\mathrm{Re}^{m}$ is the Nusselt number,
$\lambda_{g}$ is the heat conductivity of the gas, and $a_{1}$,
$a_{2}$, $m$ are numerical constants. We take $a_{1}=0$, $a_{2}=0.0425$
and $m=1$ \cite{goldshtik1984transfer}, thereby assuming purely
convective losses. The validity of this approximation for heat transfer
in flows in packed beds of solid particles is justified in \cite{goldshtik1984transfer}. 

In conservation form, the energy equation becomes 
\begin{equation}
\left(\rho e\right)_{t}+\left(\rho u\left(e+pv\right)\right)_{x}=-\frac{h}{\phi},\label{eq:energy}
\end{equation}
where 
\begin{equation}
e=e_{i}+\frac{u^{2}}{2}-\lambda Q=\frac{pv}{\gamma-1}+\frac{u^{2}}{2}-\lambda Q.
\end{equation}
The reaction rate equation is 

\begin{equation}
\lambda_{t}+u\lambda_{x}=\omega,
\end{equation}
or in conservation form, 
\begin{equation}
\left(\rho\lambda\right)_{t}+\left(\rho u\lambda\right)_{x}=\rho\omega.\label{eq:reaction}
\end{equation}

We note that these equations with appropriate modifications of the
loss terms can also be used to describe gaseous detonation in rough
tubes. The ideas that follow are expected to carry over to such a
setting without significant changes.

Across the shock-discontinuity surface, the following Rankine-Hugoniot
conditions hold: 
\begin{alignat}{1}
 & -D[\rho]+[\rho u]=0,\\
 & -D[\rho u]+[p+\rho u^{2}]=0,\\
 & -D[\rho e]+[\rho u(e+pv)]=0,\\
 & -D[\rho\lambda]+[\rho u\lambda]=0.
\end{alignat}
Here $U_{a}=\left(\rho_{a},\, u_{a},\, p_{a},\,\lambda_{a}\right)$
is the ambient state ahead of the shock, $U_{s}=\left(\rho_{s},\, u_{s},\, p_{s},\,\lambda_{s}\right)$
is the state immediately after the shock and $[j]=j_{a}-j_{s}$ denotes
a jump of $j$ across the shock. For a perfect gas, the shock conditions
can be solved explicitly and the following expressions give the post-shock
state in terms of the upstream state, 
\begin{alignat}{1}
 & \frac{p_{s}}{p_{a}}=\frac{2\gamma}{\gamma+1}M_{a}^{2}-\frac{\gamma-1}{\gamma+1},\label{eq:ps}\\
 & \frac{u_{s}}{c_{a}}=\frac{2}{\gamma+1}\frac{M_{a}^{2}-1}{M_{a}},\label{eq:us}\\
 & \frac{\rho_{s}}{\rho_{a}}=\frac{\left(\gamma+1\right)M_{a}^{2}}{2+\left(\gamma-1\right)M_{a}^{2}},\label{eq:vs}\\
 & \lambda_{s}=0.\label{eq:lambdas}
\end{alignat}
Here $M_{a}=D/c_{a}$ is the detonation Mach number with respect to
the upstream state.

\section{\label{sec:Steady-state-solution}Steady-state solutions}

In this section, we analyze the existence of the steady-state solutions
of the governing equations and calculate their properties. As always
in detonation theory, the speed of the wave is unknown \emph{a priori}
and must be determined as a solution of a nonlinear eigenvalue problem.
The goal is then to compute the steady traveling-wave structure and
its speed. In order to do that, one must carefully pose the mathematical
boundary value problem by prescribing appropriate equations and boundary
conditions. Because of the nature of the problem at hand, two apparently
distinct situations can arise, which can however be put in a unified
framework. 

In most of the literature on detonation theory, the interest is in
finding a self-sustained structure which means by definition that
the propagating shock--reaction-zone complex contains an embedded
sonic locus. This sonic locus provides acoustic confinement that precludes
any downstream influence on the detonation dynamics, thus making the
detonation wave effectively an autonomous process whose dynamics are
driven by the finite region between the shock and the sonic locus.
To compute the structure of such a wave, one requires knowledge of
the governing equations, shock conditions and appropriate conditions
at the sonic point. Knowledge of the flow conditions downstream of
the sonic locus is not required. 

Another structure that has been extensively investigated in the past
is that of overdriven detonation. This wave is obtained by providing
downstream support in the form of a piston pushing the flow in the
direction of the lead shock and thus preventing the formation of sonic
confinement. In overdriven detonation, the entire region between the
shock and the piston is subsonic. To determine the structure, knowledge
of the flow condition at the piston is necessary. This condition is
simply that the flow velocity at the piston location is the same as
the piston velocity. A boundary value problem is thus posed that requires
the construction of a smooth solution connecting the shock state with
the state at the piston. 

Thus, we arrive at a possibility that, in general, the problem of
finding the detonation structure (self-sustained or not) is posed
as follows: determine a traveling shock-wave solution of the reactive
Euler equations subject to quiescent equilibrium conditions upstream
of the shock and appropriate conditions at an infinite distance downstream
of the shock. If the detonation is self-sustained, the boundary-value
problem is divided into two sub-problems: 1) finding the structure
between the shock and the sonic point, which is independent of the
state downstream of the sonic locus, and 2) finding the flow structure
downstream of the sonic locus, which is determined by the state at
the sonic locus as well as by the conditions at an infinite distance
downstream of the shock. 

In the problem at hand, it is required to determine a traveling-wave
solution of governing equations (\ref{eq:continuity}), (\ref{eq:momentum}),
(\ref{eq:energy}) and (\ref{eq:reaction}), which consists of a lead
shock followed by a smooth reaction zone, subject to a quiescent non-reacting
state ahead of the shock and the equilibrium state in the burnt products
at infinite distance from the shock. From the governing equations,
it follows that the equilibrium state is necessarily at $\lambda=1$
and $u=0$. No other possibility exists, because the equilibrium requires
that the right-hand sides of (\ref{eq:momentum}), (\ref{eq:energy})
and (\ref{eq:reaction}) vanish at infinity, which can only happen
at $\lambda=1$ and $u=0$. The application of the Rankine-Hugoniot
conditions reduces the problem domain to a half-line between the shock
and the downstream infinity. This formulation of the boundary value
problem is general in the sense that the self-sustained solution is
part of it, but solutions without a sonic point are also admitted. 

To proceed, we rescale the governing equations with respect to the
upstream state, denoted by a subscript $a$: $\hat{\rho}=\rho/\rho_{a},$
$\hat{p}=p/p_{a}$, $u_{a}=\sqrt{p_{a}\mathrm{v}_{a}}$, $\hat{u}=u/u_{a}$,
$\hat{E}=E/u_{a}^{2},$ $\hat{Q}=Q/u_{a}^{2}$, $\hat{T}=TR/\left(p_{a}\mathrm{v}_{a}W\right)=\hat{p}/\hat{\rho}$,
$\hat{D}=D/u_{a}$. The length scale is chosen to be the half-reaction
zone length, $l_{1/2}$, of the ideal planar detonation, such that
$\hat{x}=x/l_{1/2}$. The time is rescaled as $\hat{t}=t\, u_{a}/l_{1/2}$.
In the new dimensionless variables, the governing equations retain
their form. However, the dimensionless loss terms become (dropping
the hats after the rescaling):

\begin{alignat}{1}
 & f=c_{f}\rho|u|u,\quad\quad\quad c_{f}=6b_{1}\left(1-\phi\right)\cdot\frac{l_{1/2}}{d},\\
 & h=c_{h}|u|(T-1),\quad c_{h}=\frac{9a_{2}(1-\phi)^{2}}{\phi}\cdot\frac{l_{t}l_{1/2}}{l_{v}d}.
\end{alignat}
Here, $l_{t}=\lambda_{g}W/R\sqrt{p_{a}\rho_{a}}$, $l_{v}=\nu/\sqrt{p_{a}/\rho_{a}}$
are the characteristic thermal and viscous length scales. The dimensionless
coefficients, $c_{f}$ and $c_{h}$, which are now just two numbers
measuring the effects of the momentum and heat losses, respectively,
depend on the ratios of various length scales that are characteristic
of the relevant transport processes. For $c_{f}$, it is the ratio
of the length of the reaction zone and the particle diameter. For
$c_{h}$, it is a combination of four length scales: the viscous scale,
$l_{v}$, the thermal scale, $l_{t}$, the reaction scale, $l_{1/2}$,
and the particle diameter, $d$. We note, however, that 
\begin{equation}
\frac{c_{h}}{c_{f}}=\frac{1.5a_{2}\left(1-\phi\right)}{b_{1}\phi}\,\frac{l_{t}}{l_{v}}=\frac{1.5a_{2}\left(1-\phi\right)}{b_{1}\phi}\,\frac{\lambda_{g}W}{R\nu}\label{eq:ch_over_cf}
\end{equation}
is independent of the reaction or the particle diameter, but depends
on the gas viscosity and heat conduction coefficient. We also note
that $c_{f}$ is proportional to $l_{1/2}/d$, which equals, roughly,
the number of particles that fit inside the reaction zone of the ideal
detonation. Thus, small values of $c_{f}$ correspond to particles
that are much larger than the thickness of the reaction zone in the
ideal detonation. 

The traveling-wave solution of the governing equations is sought in
the form $U=U(\xi)=\left(\rho,\, u,\, p,\,\lambda\right)(\xi)$, where$ $
$\xi=x-Dt.$ Substitution into (\ref{eq:continuity}), (\ref{eq:momentum}),
(\ref{eq:energy}) and (\ref{eq:reaction}) yields (with the primes
denoting the derivative with respect to $\xi$): 
\begin{alignat}{1}
 & \left(\rho(u-D)\right)'=0,\label{eq:mass-steady}\\
 & \left(p+\rho(u-D)u\right)'=-\frac{f}{\phi},\label{eq:momentum-steady}\\
 & \left(\rho(u-D)e+pu\right)'=-\frac{h}{\phi},\label{eq:energy-steady}\\
 & (u-D)\lambda'=\omega.\label{eq:reaction-steady}
\end{alignat}
This system of equations must be solved subject to the jump conditions
at the shock, $\xi=0$: 
\begin{alignat*}{1}
 & \rho\left(0\right)=\rho_{s}\left(D\right),\quad u\left(0\right)=u_{s}\left(D\right),\quad p\left(0\right)=p_{s}\left(D\right),\quad\lambda\left(0\right)=0,
\end{alignat*}
as given by (\ref{eq:ps}--\ref{eq:lambdas}), and the far-field conditions
at $\xi=-\infty$, which are $u=0$ and $\lambda=1$. The only other
condition, besides the requirement that the solution of the system
satisfies these boundary values, is that it be smooth and bounded
throughout the domain $-\infty<\xi\le0$. As we will see below, this
requirement is essential in order to determine the values of the detonation
speed, $D$, given all the parameters of the problem.

In the ideal case, when $f=h=0$, system (\ref{eq:mass-steady}--\ref{eq:reaction-steady})
can be integrated directly to yield the ideal detonation solution,
\begin{alignat}{1}
 & v_{0}\left(\lambda\right)=1/\rho_{0}\left(\lambda\right)=\frac{\gamma}{\gamma+1}\frac{1+D_{\CJ}^{2}}{D_{\CJ}^{2}}\left[1-\frac{D_{\CJ}^{2}-\gamma}{\gamma\left(1+D_{\CJ}^{2}\right)}\sqrt{1-\lambda}\right],\label{eq:ideal_density}\\
 & u_{0}\left(\lambda\right)=\frac{1}{\gamma+1}\frac{D_{\CJ}^{2}-\gamma}{D_{\CJ}}\left(1+\sqrt{1-\lambda}\right),\label{eq:ideal_velocity}\\
 & p_{0}\left(\lambda\right)=\frac{1+D_{\CJ}^{2}}{\gamma+1}\left[1+\frac{D_{\CJ}^{2}-\gamma}{1+D_{\CJ}^{2}}\sqrt{1-\lambda}\right],\label{eq:ideal_pressure}\\
 & \xi=\int_{0}^{\lambda}\frac{u_{0}\left(\lambda\right)-D_{\CJ}}{\omega_{0}}\der\lambda.\label{eq:ideal_lambda}
\end{alignat}
The ideal solution is chosen to be of the self-sustained nature (i.e.,
the CJ solution) with the sonic point at $\xi=-\infty$ and $\lambda=1$.
Then, 
\begin{equation}
D_{\CJ}=\sqrt{\gamma+\frac{1}{2}\left(\gamma^{2}-1\right)Q}+\sqrt{\frac{1}{2}\left(\gamma^{2}-1\right)Q}.\label{eq:D_CJ}
\end{equation}

The pre-exponential factor that imposes the half-reaction length of
unity is found from 
\begin{equation}
k=\int_{0}^{1/2}\frac{u_{0}\left(\lambda\right)-D_{\CJ}}{\left(1-\lambda\right)\exp\left(-E\rho_{0}\left(\lambda\right)/p_{0}\left(\lambda\right)\right)}\der\lambda.\label{eq:k}
\end{equation}
This value of $k$ is used in the calculations of the non-ideal detonation
when $f$ and $h$ are non-zero. In dimensional terms, if the rate
function is given as $\tilde{k}\left(1-\lambda\right)\exp\left(-\tilde{E}/\tilde{T}\right)$
(the tildes denote the dimensional quantity), then the dimensional
length of the region over which half of the reactant is burnt, is
\begin{equation}
\tilde{l}_{1/2}=\int_{0}^{1/2}\frac{\tilde{u}_{0}\left(\lambda\right)-\tilde{D}_{\CJ}}{\tilde{k}\left(1-\lambda\right)\exp\left(-\tilde{E}/\tilde{T}\right)}\der\lambda=\frac{\tilde{u}_{a}}{\tilde{k}}\int_{0}^{1/2}\frac{u_{0}\left(\lambda\right)-D_{\CJ}}{\left(1-\lambda\right)\exp\left(-E/T\right)}\der\lambda=\frac{\tilde{u}_{a}}{\tilde{k}}k.
\end{equation}
Therefore, $k=\tilde{l}_{1/2}\tilde{k}/\tilde{u}_{a}$, which identifies
the characteristic length scale in terms of the mixture properties.

The continuity equation can be integrated even in the non-ideal case
to result in 
\begin{equation}
\rho=\frac{D}{D-u}.\label{eq:rho(u)}
\end{equation}
To proceed with the solution of the remaining equations, we rewrite
(\ref{eq:mass-steady}--\ref{eq:reaction-steady}) in the matrix form:
\begin{equation}
(A-DI)U'=G,\label{eq:matrix-system}
\end{equation}
where 
\[
A=\left(\begin{array}{cccc}
u & \rho & 0 & 0\\
0 & u & 1/\rho & 0\\
0 & \gamma p & u & 0\\
0 & 0 & 0 & u
\end{array}\right),\; G=\left(\begin{array}{c}
0\\
F\\
H\\
\omega
\end{array}\right),
\]
$I$ is the unit matrix, and 
\begin{equation}
F=-\frac{f}{\rho\phi},\quad H=(\gamma-1)Q\rho\omega+(\gamma-1)\left(\frac{uf-h}{\phi}\right).
\end{equation}
The acoustic eigenvalues of matrix $A$ are $s_{1}=u+c$ and $s_{2}=u-c$,
where $c=\sqrt{\gamma pv}$ is the sound speed. Their corresponding
left eigenvectors are $l_{1}=\left(0,\,1,\, c/\gamma p,\,0\right)$
and $l_{2}=\left(0,\,1,\,-c/\gamma p,\,0\right)$. Left-multiplying
(\ref{eq:matrix-system}) by $l_{1}$, we obtain,
\begin{equation}
(s_{1}-D)(u'+\frac{c}{\gamma p}p')=F+\frac{c}{\gamma p}H.\label{eq:s1_char_eqn}
\end{equation}
Similarly, left-multiplying (\ref{eq:matrix-system}) by $l_{2}$,
we obtain,
\begin{equation}
(s_{2}-D)(u'-\frac{c}{\gamma p}p')=F-\frac{c}{\gamma p}H.\label{eq:s2_char_eqn}
\end{equation}
From (\ref{eq:s1_char_eqn}, \ref{eq:s2_char_eqn}), it follows that
\begin{alignat}{1}
 & \frac{\der u}{\der\xi}=\frac{1}{2}\left[\frac{F+\frac{c}{\gamma p}H}{u+c-D}+\frac{F-\frac{c}{\gamma p}H}{u-c-D}\right],\label{eq:u_eqn}\\
 & \frac{\der p}{\der\xi}=\frac{\gamma p}{2c}\left[\frac{F+\frac{c}{\gamma p}H}{u+c-D}-\frac{F-\frac{c}{\gamma p}H}{u-c-D}\right].\label{eq:p_eqn}
\end{alignat}

The importance of this form of the equations is that it makes evident
that the solution is not necessarily regular if the denominators of
(\ref{eq:u_eqn}) or (\ref{eq:p_eqn}) vanish somewhere at $\xi<0$.
Note that at $\xi=0$, they cannot vanish due to the Lax conditions.
Because $u-D$ is the flow velocity relative to the shock, which must
be negative for the shock propagating to the right, it follows that
$u-c-D$ can never vanish. However, $u+c-D$ can vanish. If it does,
we must require that at the same point (which is the sonic point)
$F+cH/\gamma p=0$ as well. This is basically the statement of the
generalized CJ condition \cite{Eyring49,WoodKirkwood54} and this
condition is necessary to determine the value of $D$ when the flow
contains a sonic point. It must be emphasized that the existence of
the sonic point is not necessary if one can find a smooth solution
that connects the shock state with the far-field equilibrium conditions.
Irrespective of the presence of the sonic point, the fundamental problem
is that of finding a smooth and bounded solution of the boundary value
problem at hand. 

Another important consequence of (\ref{eq:u_eqn}--\ref{eq:p_eqn})
is that the only possible equilibrium solution at $\xi=-\infty$ is
that with $u=0$. Indeed, setting the right-hand sides of both equations
to zero, we obtain that in equilibrium both $F$ and $H$ must vanish.
Therefore, from $F=0$, we obtain $u=0$ and, as a consequence of
this and of $\lambda=1$, it also follows that $H=0$. No other possibility
exists. 

If a sonic locus denoted by $\xi_{*}$ exists, then at that point
by definition, $u_{*}+c_{*}-D=0.$ Requiring that $F+cH/\gamma p$
vanish at $\xi_{*}$ as well, one obtains the generalized CJ conditions,
\begin{alignat}{1}
 & \frac{c_{*}}{\gamma p_{*}}(\gamma-1)\left(Q\rho_{*}\omega(\lambda_{*})+\frac{u_{*}f_{*}-h_{*}}{\phi}\right)-\frac{f_{*}}{\rho_{*}\phi}=0,\label{eq:thermicity}\\
 & u_{*}+c_{*}-D=0.\label{eq:sonicity}
\end{alignat}
This is a system of equations for $D$ and $\lambda_{*}$ that in
principle allows one to find the full solution. However, in practice
it is extremely difficult to solve numerically. To illustrate the
required calculations, we assume that all parameters of the problem
are given. Then, the only unknowns are $D$ and the solution profiles. 

To remind the reader, in order to determine $D$ in the ideal case,
one finds the full solution as a function of $\lambda$ as $p=p\left(\lambda,D\right)$,
$u=u\left(\lambda,D\right)$, and $\rho=\rho\left(\lambda,D\right)$
(given by (\ref{eq:ideal_density}--\ref{eq:ideal_lambda})), which
depend parametrically on $D$ ($\lambda$ as a function of $\xi$
is found implicitly by integrating (\ref{eq:ideal_lambda})). The
solution is then substituted into the CJ conditions (\ref{eq:thermicity}--\ref{eq:sonicity}).
These latter in the ideal case are simply $\omega_{*}=0$ and $u_{*}+c_{*}-D=0$;
therefore, $\lambda_{*}=1$ and the second condition yields the desired
result for $D$, as given by (\ref{eq:D_CJ}).

In the non-ideal case, however, the solution profiles are unavailable
analytically and must be determined numerically. The procedure of
finding $D$ is in principle, to guess the value of $D$, integrate
the ordinary differential equations (ODE) for $p$ and $u$ in the
$\lambda$-variable from the shock until $\lambda_{*}$ such that
(\ref{eq:thermicity}--\ref{eq:sonicity}) are satisfied. This is
done iteratively by varying $D$ until the sonic conditions are satisfied.
Generally, the sonic point has a saddle character and, therefore,
the integration process just described is extremely ill-conditioned.
For this reason, it is often preferable first to identify the sonic
locus, then linearize the governing ODE system in its neighborhood,
step out analytically by a small step and then integrate numerically
back to the shock. This procedure is better conditioned and has been
used in the past. Yet, it is not a good choice for our problem either.
One reason is that the sonic locus is unknown \emph{a priori, }so
that linearization about it does not eliminate the algorithmic complexity
of the search for $D$ and the locus itself. The second and more serious
reason is that the method fails when the solution contains no sonic
locus. The latter is a possibility that cannot be ignored as one does
not in general know \emph{a priori }whether or not a sonic point exists
in the solution.

To resolve the difficulties mentioned in the previous paragraph, we
suggest a change of dependent variables that completely eliminates
the singular behaviour from the governing equations. Note that (\ref{eq:s1_char_eqn}--\ref{eq:s2_char_eqn})
can be written as 
\begin{alignat}{1}
 & \frac{\der u}{\der\xi}+\frac{c}{\gamma p}\frac{\der p}{\der\xi}=\frac{F+\frac{c}{\gamma p}H}{u+c-D},\label{eq:u_eqn-1}\\
 & \frac{\der u}{\der\xi}-\frac{c}{\gamma p}\frac{\der p}{\der\xi}=\frac{F-\frac{c}{\gamma p}H}{u-c-D}.\label{eq:p_eqn-1}
\end{alignat}
In this system, only the first equation contains a potential singularity.
Note now that 
\[
\frac{c}{\gamma p}=\frac{\sqrt{\gamma p/\rho}}{\gamma p}=\sqrt{\frac{D-u}{D\gamma p}}.
\]
With this substitution, the left-hand sides of (\ref{eq:u_eqn-1},\ref{eq:p_eqn-1})
become 
\[
2\sqrt{D-u}\frac{\der}{\der\xi}\left(-\sqrt{D-u}\pm\sqrt{\frac{p}{\gamma D}}\right),
\]
which motivates the introduction of new variables,
\begin{alignat}{1}
 & r=-\sqrt{D-u}-\sqrt{\frac{p}{\gamma D}},\label{eq:r}\\
 & s=-\sqrt{D-u}+\sqrt{\frac{p}{\gamma D}}.\label{eq:s}
\end{alignat}
Going back to $u$ and $p$ in terms of $r$ and $s$ is easy because
$r+s=-2\sqrt{D-u}$ and $r-s=-2\sqrt{p/\gamma D}$. It is interesting
to compute $u+c-D$ in terms of the new variables. The speed of sound
is 
\[
c=\sqrt{\frac{\gamma p}{\rho}}=\gamma\sqrt{\frac{p}{\gamma D}\left(D-u\right)}=\gamma\frac{r^{2}-s^{2}}{4},
\]
where we took into consideration that $\left|r\right|>\left|s\right|$
at any $p>0$. Therefore, 
\begin{equation}
u+c-D=-\frac{\left(r+s\right)^{2}}{4}+\gamma\frac{r^{2}-s^{2}}{4}=\frac{\gamma-1}{4}\left(r+s\right)\left(r-\mu s\right),
\end{equation}
where 
\begin{equation}
\mu=\frac{\gamma+1}{\gamma-1}.\label{eq:mu}
\end{equation}
Thus the sonic condition $u_{*}+c_{*}-D=0$ is reduced to a simple
linear equation in terms of the new variables, $r$ and $s$,
\begin{equation}
r_{*}-\mu s_{*}=0.\label{eq:rs_sonic_condn}
\end{equation}

Equations (\ref{eq:u_eqn}--\ref{eq:p_eqn}) lead to a new system
of ODE for $r$, $s$ and $\lambda$, 
\begin{alignat}{1}
 & \frac{\der r}{\der\xi}=\frac{\gamma-1}{2\rho c\sqrt{D-u}\left(u-c-D\right)}\left[-Q\rho\omega+\frac{h}{\phi}-\left(u+\frac{c}{\gamma-1}\right)\frac{f}{\phi}\right]\equiv G(r,s,\lambda),\label{eq:r_ode}\\
 & \frac{\der s}{\der\xi}=-\frac{1}{\rho c\left(D-u\right)\left(r-\mu s\right)}\left[-Q\rho\omega+\frac{h}{\phi}-\left(u-\frac{c}{\gamma-1}\right)\frac{f}{\phi}\right]\equiv\frac{g(r,s,\lambda)}{r-\mu s},\label{eq:s_ode}\\
 & \frac{\der\lambda}{\der\xi}=\frac{\omega}{u-D}\equiv L(r,s,\lambda).\label{eq:lambda_ode}
\end{alignat}
Singularity is contained only in (\ref{eq:s_ode}), but it is now
of a very simple form in the new variables. By introducing another
change of variables, we can completely remove the singularity from
the system. Let
\begin{equation}
z=(r-\mu s)^{2},\label{eq:z}
\end{equation}
to replace one of the variables, say 
\begin{equation}
s=\frac{1}{\mu}\left(r\pm\sqrt{z}\right).\label{eq:s_of_z}
\end{equation}
Before proceeding, it is important to note that the inversion of the
variables from $r,z$ to $r,s$ is \emph{not} single-valued. The change
from one branch of the transformation to another happens exactly at
the sonic point, which is now located at $z=0$, where the Jacobian
of the transformation vanishes. The positive sign in (\ref{eq:s_of_z})
corresponds to the subsonic flow in the reference frame of the shock,
while the negative sign corresponds to the supersonic flow. Indeed,
at the shock $u+c-D=\left(\gamma-1\right)\left(r+s\right)\left(r-\mu s\right)/4>0$
and since $r+s=-2\sqrt{D-u}<0$, then $r-\mu s$ should be $<0$ in
the subsonic region between the shock and the sonic point. This amounts
to selecting the positive sign in (\ref{eq:s_of_z}). Downstream of
the sonic point, the sign switches to negative. 

By removing the $s$ variable from (\ref{eq:r_ode}--\ref{eq:lambda_ode})
in favor of $z$, we obtain a new system: 
\begin{alignat}{1}
 & \frac{\der r}{\der\xi}=G_{\pm}(r,z,\lambda),\label{eq:r_ode_new}\\
 & \frac{\der z}{\der\xi}=\pm2\sqrt{z}G_{\pm}(r,z,\lambda)-2\mu g_{\pm}(r,z,\lambda),\label{eq:z_ode_new}\\
 & \frac{\der\lambda}{\der\xi}=L_{\pm}(r,z,\lambda),\label{eq:lambda_ode_new}
\end{alignat}
which is now completely free of singularity. Here, $G_{\pm}$ and
$g_{\pm}$ are the same functions as $G$ and $g$ in (\ref{eq:r_ode}--\ref{eq:s_ode})
with the $\pm$ sign indicating the branch chosen in (\ref{eq:s_of_z}).
Of course, the transonic character of the solutions cannot just disappear
from the equations. There is deep physics behind the existence of
the sonic point in the solution that must remain somehow in the equations.
It indeed does remain in the form of a switch from one branch of the
square root to another when passing through the sonic point. 

Equations (\ref{eq:r_ode_new}--\ref{eq:lambda_ode_new}) must be
solved subject to the shock conditions, 
\begin{alignat}{1}
 & r\left(0\right)=r_{s},\label{eq:r(0)}\\
 & z(0)=\left(r_{s}-\mu s_{s}\right)^{2},\label{eq:z(0)}\\
 & \lambda(0)=0,\label{eq:lambda(0)}
\end{alignat}
where 
\begin{alignat}{1}
 & r_{s}=-\sqrt{D-u_{s}}-\sqrt{\frac{p_{s}}{\gamma D}},\label{eq:rs}\\
 & s_{s}=-\sqrt{D-u_{s}}+\sqrt{\frac{p_{s}}{\gamma D}}\label{eq:ss}
\end{alignat}
and $u=0$ at $\xi=-\infty$, which amounts to 

\begin{alignat}{1}
 & r\left(-\infty\right)=-\sqrt{D}-\sqrt{\frac{p\left(-\infty\right)}{\gamma D}},\label{eq:r_infty}\\
 & z\left(-\infty\right)=-\sqrt{D}+\sqrt{\frac{p\left(-\infty\right)}{\gamma D}}.\label{eq:z_infty}
\end{alignat}
As was mentioned previously, the cases where the solution contains
no sonic point and therefore the entire flow downstream of the shock
remains subsonic should not be excluded from consideration. Condition
$u\left(-\infty\right)=0$ leads to conditions on $r$ and $z$ given
by (\ref{eq:r_infty}--\ref{eq:z_infty}). However, the latter depend
on $p\left(-\infty\right)$, which cannot be enforced, in general.
If the solution contains a sonic point, then the regularization conditions
(\ref{eq:thermicity}--\ref{eq:sonicity}) become simply 
\begin{equation}
\frac{\der z}{\der\xi}=0\;{\mbox{at }}z=0.\label{eq:new_sonic_condn}
\end{equation}
In this case, the equilibrium conditions at $\xi=-\infty$ must still
be enforced, but they are now required only for the determination
of the solution downstream of the sonic locus. If the latter does
not exist, the system (\ref{eq:r_ode_new}--\ref{eq:lambda_ode_new})
must be solved subject to three conditions at the shock, (\ref{eq:r(0)}--\ref{eq:lambda(0)}),
and the conditions at $\xi=-\infty$. This set of boundary conditions
is expected to be complete on physical grounds in order to yield a
well-defined solution of the problem. The solution is not necessarily
unique, as in previous studies of closely related problems. A typical
result is a multi-valued solution for the detonation speed, $D$,
as a function of, for example, the friction coefficient, $c_{f}$,
given that all the other parameters are fixed. However, in contrast
to the previous work, we find that the boundary-value problem as posed
above yields \emph{set-valued} solutions at certain parameters. In
other words, a continuos range of solutions with $D$ varying in some
interval can exist for a given mixture and given loss conditions.
Thus, the nonlinear eigenvalue problem at hand can have not only a
discrete but also a continuous spectrum.

\section{\label{sec:Example-calculation}Numerical results}

\subsection{The integration algorithm}

The change of dependent variables to $r$ and $z$ is very helpful
as it eliminates the numerical difficulties that are inevitable in
the integration of the original system (\ref{eq:r_ode}-\ref{eq:lambda_ode}).
These difficulties have previously been avoided in simpler situations
(for example, when the governing equations can be reduced to a single
ODE), by integrating from the sonic point to the shock \cite{zel1987detonation,higgins2012steady,bdzil2012theory}.
This algorithm works well if the sonic point location is known \emph{a
priori}. If, however, the location of the sonic point depends on the
shock speed and other parameters, which one is attempting to find,
then the algorithm becomes unwieldy, even though still valid. 

\begin{figure}[H]
\noindent \begin{centering}
\includegraphics[clip,width=12cm]{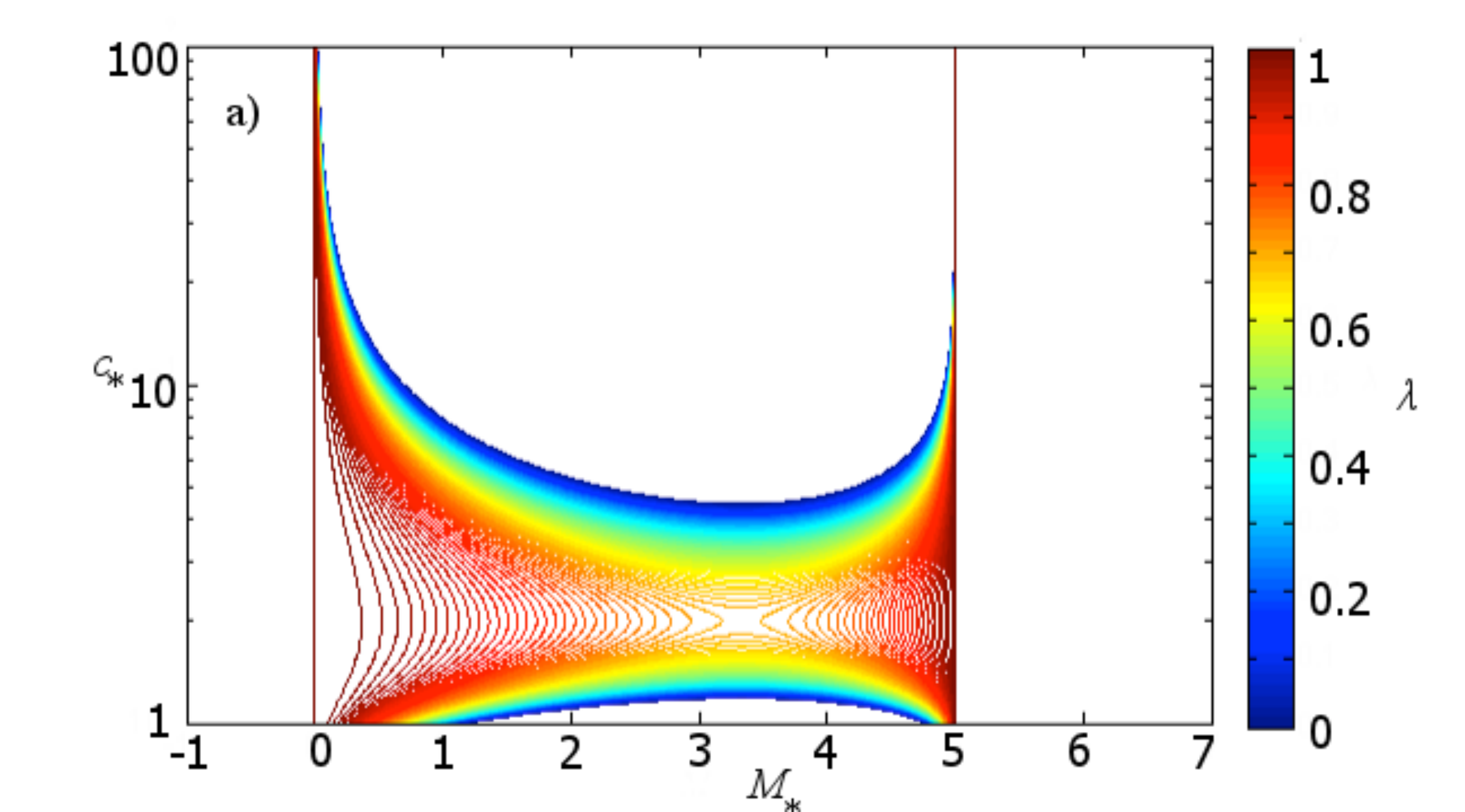}
\par\end{centering}

\noindent \begin{centering}
\includegraphics[clip,width=12cm]{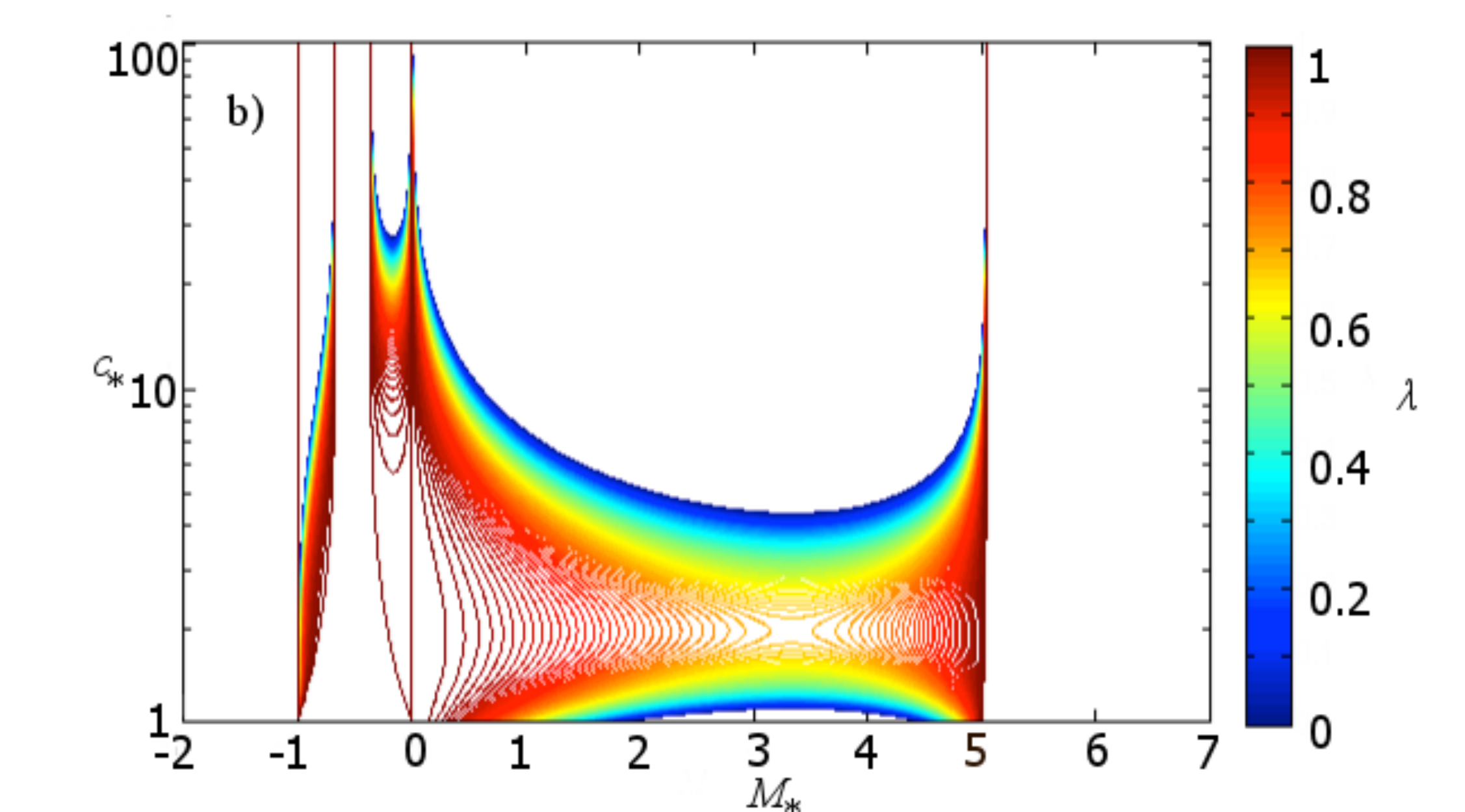}
\par\end{centering}

\caption{\label{fig:sonic-locus-location regions} The regions in the $M_{*}$-$c_{*}$
space where the sonic locus can exist: a) in the presence of only
friction losses, \textbf{$c_{f}=0.01$}, $c_{h}=0$, and b) when both
heat and friction losses are present,\textbf{ $c_{f}=0.01$}, $c_{h}=1.5c_{f}$.
The lines in the figures are the iso-lines of $\lambda$ taken with
step $0.01$. }
\end{figure}

Figure \ref{fig:sonic-locus-location regions} illustrates the difficulty.
If we choose to integrate from the sonic locus, the algorithm proceeds
as follows. Assume that all the parameters of the problem are given
such that we need to determine $D$ and the solution profiles. First,
we need to identify the sonic locus. Equations (\ref{eq:thermicity}--\ref{eq:sonicity})
must hold at the sonic point. Given that $\rho=D/\left(D-u\right)$,
$p=\rho c^{2}/\gamma$ and $T=p/\rho$, these two equations can be
written in terms of $u_{*}$, $c_{*}$, $\lambda_{*}$ and $D$. One
can eliminate, say, $D$ using (\ref{eq:sonicity}) and obtain one
equation in terms of $u_{*}$, $c_{*}$ and $\lambda_{*}$ by substituting
$D=u_{*}+c_{*}$ into (\ref{eq:thermicity}). To proceed, one has
to make a guess on \emph{two} sonic variables and then find the third
one by solving (\ref{eq:thermicity}). Only then would the sonic state
be known completely, after which one can proceed as usual with a numerical
integration out of the sonic point back to the shock, check the shock
conditions, and, if they are not satisfied, iterate on the guesses
of the two sonic state variables. 

The main difficulty in the procedure just described is the need to
iterate on two sonic variables as opposed to one in simpler situations.
In order to give an idea of the range of the variables that must be
guessed, we plot the domains in the plane of $M_{*}=u_{*}/c_{*}$
and $c_{*}$, the regions where (\ref{eq:thermicity}--\ref{eq:sonicity})
has a solution. The shaded areas are the regions bounded by $\lambda_{*}=0$
and $\lambda_{*}=1$ in (\ref{eq:thermicity}--\ref{eq:sonicity}).
The search algorithm for the sonic locus can, in principle, yield
the solution anywhere within the shaded areas. The most striking property
of the regions is the presence of highly intricate features, especially
at negative velocities, in the case when both heat and friction losses
are present (Fig. \ref{fig:sonic-locus-location regions}b)). When
only momentum loss is present, negative velocity at the sonic point
is impossible (Fig. \ref{fig:sonic-locus-location regions}(a)). 

In terms of our new variables, the sonic locus is fixed and determined
exactly by $z=z'=0$ at any parameters. Moreover, the singularity
is no longer present in the system. More precisely, the singularity
is reflected by the requirement that the branch of the square root
of $z$ must change as one crosses the sonic locus. This procedure
is numerically benign. Thus, we solve (\ref{eq:r_ode_new}-\ref{eq:lambda_ode_new})
with the corresponding boundary and sonic conditions, provided the
sonic point exists. The non-singular character of the system allows
one to calculate the solution with high precision. The goal is to
find such a $D$ that the solution of equations (\ref{eq:r_ode_new}-\ref{eq:lambda_ode_new})
satisfies the boundary conditions at the shock and at $\xi=-\infty$
and to study the dependence of the detonation speed on the loss factors
$c_{f}$ and $c_{h}$ assuming that other parameters are fixed. If
there is a sonic point, we require that the sonic conditions $z=z'=0$
are satisfied within machine precision. If there is no sonic point,
the condition at the left boundary of the domain is imposed that both
$u$ and $1-\lambda$ are close to zero within $10^{-10}$. Any changes
in $c_{f}$ within the machine precision result in the changes of
$u$ and $1-\lambda$ at infinity within $10^{-10}$, so that requiring
even smaller errors is not feasible. 

The new numerical algorithm works as follows. We first fix $E,$ $Q,$
$\gamma$ and the ratio $c_{f}/c_{h}$. The latter is essentially
a constant for a given mixture, as seen from (\ref{eq:ch_over_cf}).
Then, for every given $D\in[c_{a},D_{\CJ}]$, we look for the $c_{f}$
that provides a smooth solution to our boundary value problem. Here
$c_{a}$, the sound speed in the ambient state, is equal to $\sqrt{\gamma}$
in our scales. The calculations show that there are two possible cases.
If the detonation speed is less than $D_{\CJ}$, but greater than
some value $D_{s}$, which is constant for the given set of parameters,
then there is a sonic point inside the reaction zone. On the other
hand, if $D\in(c_{a},D_{s})$, then there exist smooth solutions that
satisfy the boundary conditions and remain subsonic from the shock
down to $\xi=-\infty$. 

\begin{figure}[H]
\noindent \begin{centering}
\includegraphics[clip,width=4.5cm]{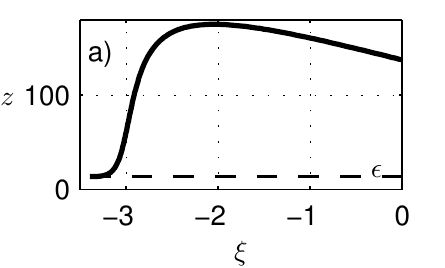}\includegraphics[width=4.5cm]{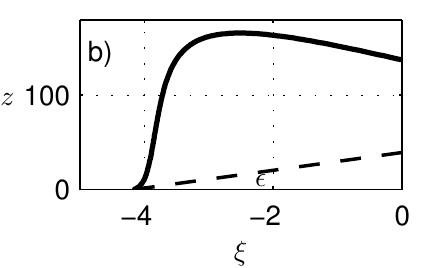}\includegraphics[width=4.5cm]{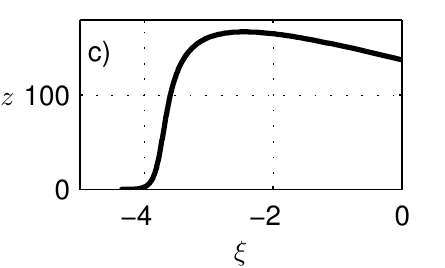}
\par\end{centering}

\caption{\label{fig:z profiles near the sonic point} Profiles of $z$ in the
shooting method for $D=0.8D_{\CJ}$ and various $c_{f}$: a) - $c_{f}=0.01$;
b) - $c_{f}=0.006$; c) - $c_{f}=0.007$. Here $E=30,\; Q=20,\;\gamma=1.2$
and $c_{h}=0$. }
\end{figure}

The possible outcomes of the numerical search for $c_{f}$ for a given
$D$ are shown in Fig. \ref{fig:z profiles near the sonic point}.
The figures are computed by shooting from the shock to the sonic point
with various $c_{f}.$ The Matlab \emph{ode15s} solver with a reference
tolerance value of $10^{-13}$ is used. In Fig. \ref{fig:z profiles near the sonic point}a),
one can see the profile of $z$ for $c_{f}$ that is too large. Here,
$z$ reaches its minimum value, but this value is positive. At $\xi$
less than the minimum point, $z$ together with $u$ and $p$ begin
to grow without bound. For the exact solution, we expect $z$ to vanish
at the minimum point. Hence, to measure the closeness of the computed
profile to the exact solution, we define the error function, $\epsilon$,
to be the minimum value of $z$. On the other hand, if $c_{f}$ is
too small, then $z$ will reach zero with a positive slope, as in
Fig. \ref{fig:z profiles near the sonic point}b). This means that
condition (\ref{eq:new_sonic_condn}) is not satisfied and $u$ will
blow up because of the infinite derivative at the sonic point. In
this case, we define $\epsilon$ to equal the derivative of $z$ at
the sonic point. Then we find the minimum value of $\epsilon(c_{f})$
using Matlab's \emph{FminSearch} operator with a tolerance value of
$10^{-13}$. The minimum value of $\epsilon=0$ corresponds to Fig.
\ref{fig:z profiles near the sonic point}c). This case is the one
that satisfies condition (\ref{eq:new_sonic_condn}) and the corresponding
$c_{f}$ is the proper value for the given detonation speed. At this
value of $c_{f}$, we switch to the second branch of $\sqrt{z}$ and
continue the calculations on the other side of the sonic point until
$\xi=-\infty$.

\begin{figure}[H]
\noindent \begin{centering}
\includegraphics[clip,height=4.43cm]{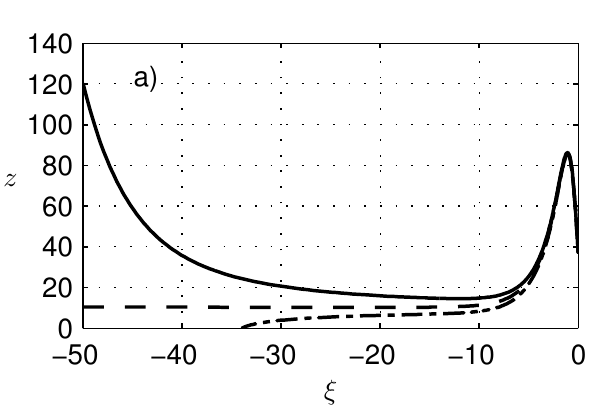}\includegraphics[clip,height=4.3cm]{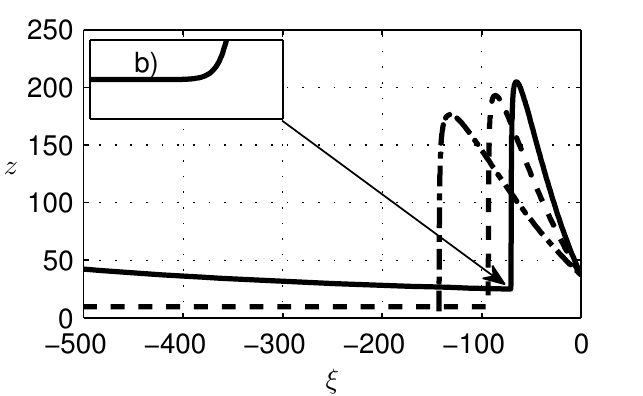}
\par\end{centering}

\caption{\label{fig:z profiles for low detonation speed} Profiles of $z$
in the shooting process at $D=0.4D_{\CJ}$, $c_{h}=0$, $Q=20$, $\gamma=1.2$
and various $c_{f}$: $c_{f}=0.01$ (solid), $c_{f}=0.0022$ (dash),
$c_{f}=0.0011$ (dash-dot); a) $E=5$ and b) $E=30$.}
\end{figure}

For $D<D_{s}$, such numerical algorithm shows that there exists no
$c_{f}$ that gives $\epsilon(c_{f})=0.$ However, smooth and bounded
subsonic solutions still exist, i.e. $z$ remains positive in $(-\infty,0)$
and $u(-\infty)=0$. Such solutions are investigated below. The shooting
from the shock gives three possibilities, as shown in Fig. \ref{fig:z profiles for low detonation speed}a).
If $c_{f}$ is too large, then $z$ passes its minimum point and then
grows to infinity (solid line). If $c_{f}$ is too small, then again
$z$ reaches the zero value with a non-zero slope, which means the
blow-up of velocity (dash-dot line). However, instead of the behaviour
in Fig. \ref{fig:z profiles near the sonic point}c), here we obtain
the dashed line in the middle, which is the solution with a bounded
value of $z$ at $\xi=-\infty$. This solution is possible only in
the case when $u(-\infty)=0$, since it is the only equilibrium state
for equations (\ref{eq:r_ode_new}-\ref{eq:lambda_ode_new}). It is
important to note that the kinks in the curves in Fig. \ref{fig:z profiles for low detonation speed}b)
indicate a smooth, but rapid change in the solution. It occurs due
to the exponentially sensitive terms in the governing equations. 

\subsection{Solutions in the presence of both heat and friction losses}

\begin{figure}[H]
\noindent \begin{centering}
\includegraphics[clip,height=4.43cm]{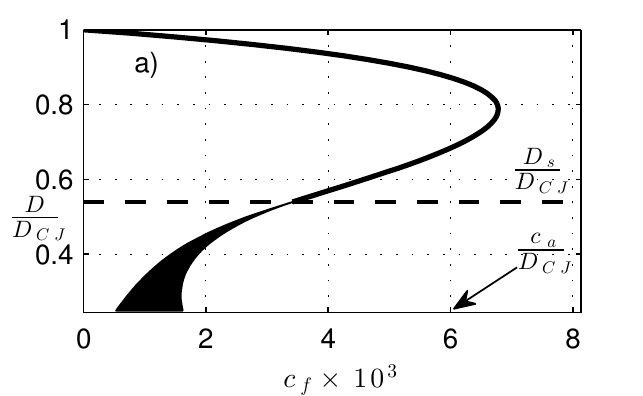}\includegraphics[clip,height=4.43cm]{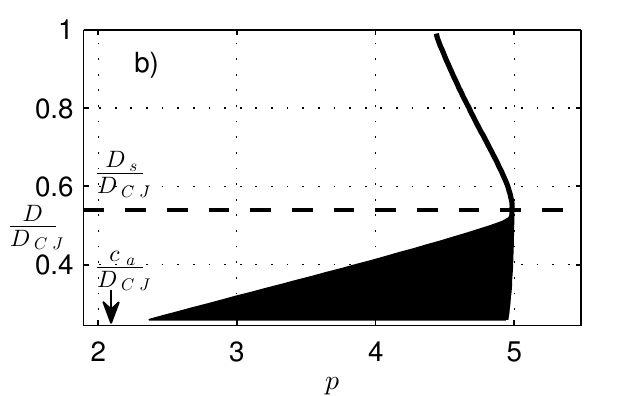}
\par\end{centering}

\caption{\label{fig:D-cf-and-D-p at infty} a) The dependence of the detonation
speed, $D$, on the loss factor, $c_{f}$; b) the pressure at infinity,
$p\left(-\infty\right)$, for the $D$-$c_{f}$ dependence shown in
a). In both cases $E=30$.}
\end{figure}

If $D_{s}<D<D_{\CJ}$, the steady-state solution passes through a
sonic point, where conditions (\ref{eq:new_sonic_condn}) must be
satisfied. Imposition of these conditions yields the exact value of
the detonation speed for the given loss terms. The presence of a sonic
point is a special feature that constrains the solutions to a particular
form, essentially independent of the boundary conditions at infinity.
However, if $c_{a}<D<D_{s}$, the sonic point and hence the constraint
disappears and the nature of the solutions changes, since now the
downstream boundary condition determines the detonation speed \cite{brailovsky2000hydraulic-a,brailovsky2002effects}.
The solutions connect the Rankine-Hugoniot conditions at the shock,
$\xi=0$, and some appropriate conditions at $\xi=-\infty$. In the
reference frame of the shock, the flow remains subsonic throughout.
As a consequence of the loss of the sonic constraints, which if present
pin down the solution to yield a unique $c_{f}$ for a given $D$,
a degree of freedom is gained. The only condition at $\xi=-\infty$
that can justifiably be imposed is $u=0$, which leaves one condition
there to be free. This can either be pressure or temperature; the
density at infinity is the same as in the ambient gas, which follows
easily from the continuity equation (\ref{eq:rho(u)}). As a result
of this degree of freedom, the case $c_{a}<D<D_{s}$ yields a set-valued
solution in which the pressure at $\xi=-\infty$ is allowed to vary
in some range. That is, given the shock conditions and $u\left(-\infty\right)=0$,
for any chosen $D\in\left(c_{a},D_{s}\right)$, there exists a continuous
range of $c_{f}$ (at fixed $c_{f}/c_{h}$) that yields a smooth and
bounded solution of (\ref{eq:r_ode_new}-\ref{eq:lambda_ode_new}).
The transition from the single-valued to set-valued solutions is shown
in Fig. \ref{fig:D-cf-and-D-p at infty}a). For the existence of the
set-valued solutions, the pressure at infinity cannot simply be arbitrary.
For a fixed $D<D_{s}$, as we vary $c_{f}$, the flow variables $p$
and $T$ at $\xi=-\infty$ vary as well (but $u=0$ and $\lambda=1$
remain the same). The range of the corresponding variation of $p\left(-\infty\right)$
is shown in Fig. \ref{fig:D-cf-and-D-p at infty}b). It is important
to keep in mind that in Fig. \ref{fig:D-cf-and-D-p at infty}b), $c_{f}$
is not fixed, but rather varies in the range seen in Fig. \ref{fig:D-cf-and-D-p at infty}a)
for any given $D$. An interesting feature of the pressure at infinity
is that its maximum value appears to coincide with the moment when
the sonic locus is exactly at infinity. This is the case in all of
the calculations that we have performed. As the velocity decreases
further, the pressure at infinity (hence the temperature there as
well) begins to drop. Therefore, the numerical calculations indicate
that the onset of the set-valued solutions coincides with the maximum
temperature (or pressure) in the products. 

In Fig. \ref{fig:D-cf-and-D-p at infty} and all of the calculations
below, unless otherwise indicated, the following parameters are fixed:
$Q=20,\;\gamma=1.2$ and $c_{h}=0.4c_{f}$.

For completeness, we analyze the structure of the solutions on the
function-valued branch where they correspond to the self-sustained
detonations. One of the main problems in the analysis is to understand
the relative role of the three driving physical processes: chemical
reaction, friction and heat transfer. While the roles of the chemical
reaction and heat transfer are relatively simple, the former always
contributing energy and the latter always taking it away, the role
of friction is more complex. The friction term, $f$, changes its
sign with velocity, which can result in momentum gain as opposed to
momentum loss when $u<0$. More importantly, however, the friction
results in heating of the gas and thus acts as a heat source. This
is seen in the energy equation (\ref{eq:energy_p}). Here, we analyze
the role of these terms in more detail by evaluating their contributions
to the fluid acceleration, $\der u/\der\xi$, and to the temperature
gradient, $\der T/\der\xi$. 

If one writes out the equation for $u\left(\xi\right)$ from (\ref{eq:u_eqn})
as follows 
\begin{equation}
\frac{\der u}{\der\xi}=\frac{1}{\left(u-D\right)^{2}-c^{2}}\left[-\left(\gamma-1\right)Q\omega-\frac{\left(\gamma u-D\right)}{\rho\phi}f+\frac{\left(\gamma-1\right)}{\rho\phi}h\right],\label{eq:u-eqn-detailed}
\end{equation}
then it is clear that the behaviour of $u$ depends on the competition
between the heat release, friction and heat loss. Recall that $f=c_{f}\rho u\left|u\right|$,
$h=c_{h}\left|u\right|\left(T-1\right)$ and that $c_{h}=\beta c_{f}$
for some fixed $\beta$ ($\beta=0.4$ in most of our calculations). 

It is interesting that $\gamma u-D$ in the friction term in (\ref{eq:u-eqn-detailed})
can take either sign. To evaluate it at the shock, we use the jump
condition (\ref{eq:us}), 
\begin{equation}
\gamma u_{s}-D=c_{a}\left(\gamma\frac{u_{s}}{c_{a}}-M_{a}\right)=c_{a}\left[\frac{\left(\gamma-1\right)M_{a}^{2}-2\gamma}{\left(\gamma+1\right)M_{a}}\right].
\end{equation}
Therefore, $\gamma u_{s}-D<0$ if $M_{a}^{2}<2\gamma/\left(\gamma-1\right)$
and $\gamma u_{s}-D>0$ otherwise. In the tail of the reaction zone,
where $u$ is small, $\gamma u-D<0$. Thus, for sufficiently strong
detonations, the factor $\gamma u-D$ will change signs somewhere
in the reaction zone. For sufficiently large $M_{a}$, we obtain that
near the shock $\left(\gamma u-D\right)f>0$ and, therefore, since
$\left(u-D\right)^{2}-c^{2}<0$ near the shock, the friction term
leads to a velocity decrease as does the heat release term. The effect
of the heat loss term is the opposite. At small enough detonation
Mach numbers, however, the role of the friction term is reversed because
the sign of $\gamma u-D$ changes near the shock. The reversal of
the sign of $\gamma u-D$ seems always to occur well above the turning
point of the $D-c_{f}$ curve and hence no significant qualitative
change to the solution results. However, we have not explored the
whole range of possibilities and the consequences of this sign change.
In the cases we considered, there seems to be no significant effect.

In the set-valued part of the solution, the flow is fully subsonic.
Therefore, $\left(u-D\right)^{2}-c^{2}<0$ everywhere. If the numerator
on the right-hand side of (\ref{eq:u-eqn-detailed}) is also negative,
then $u'>0$, such that $u$ decreases as $\xi$ tends to $-\infty$.
Suppose now that $u$ reaches its zero value. At that point, both
$f$ and $h$ vanish, but $\omega$ is still non-zero. Therefore,
$u$ will continue to decrease. Once $u$ becomes negative, the $h$
term retains its sign, i.e., it remains positive, but the $f$ term
now switches sign. Instead of acting with the reaction term, which
it did for $u>0$, now the friction term acts with the heat loss term.
As the magnitude of the negative velocity keeps increasing, the friction
and heat loss terms in the numerator of (\ref{eq:u-eqn-detailed})
will balance the diminishing reaction rate term at some point. Downstream
of that point, it is the loss terms that dominate, while the reaction
term stops playing any significant role. Since the sign of $u'$ is
now reversed, $u$ begins to increase toward zero. The trend is asymptotic,
since at small $u$, the right-hand side of (\ref{eq:u-eqn-detailed})
is essentially proportional to $-\left|u\right|$, such that $u=0$
near infinity is a (half-) stable fixed point. 

The equation for temperature is more revealing. Using $T'=p'v+pv'$,
$v=\left(D-u\right)/D$ and (\ref{eq:u_eqn}-\ref{eq:p_eqn}) to calculate
$p'$ and $u'$, we find that 
\begin{alignat}{1}
 & \frac{\der T}{\der\xi}=-\frac{\left(\gamma-1\right)Dv}{\gamma\left(1-M^{2}\right)M^{2}c^{2}}\left[\left(1-\gamma M^{2}\right)\left(Q\omega-\frac{h}{\rho\phi}\right)+\left(1-\frac{\gamma u}{D}M^{2}\right)\frac{Df}{\rho\phi}\right].\label{eq:T-eqn-detailed}
\end{alignat}
Here, $M=\left(D-u\right)/c$ is the local Mach number of the flow.
Because the factor in front of the brackets in (\ref{eq:T-eqn-detailed})
is always negative, the rise of the temperature in the reaction zone
is due to the positive terms in the brackets. Depending on whether
$M^{2}$ is smaller or larger than $1/\gamma$, the heat release term
is seen to contribute to the temperature increase or decrease, respectively.
The latter happens when the cooling by expansion dominates over heating,
a situation well known in ideal detonation theory. This dual role
of heat release leads to a temperature maximum in the reaction zone.
The heat loss term is seen to have an effect opposite to that of the
heat release. Note that the roles of chemical heat release and the
heat transfer are always opposite and they switch signs where $1-\gamma M^{2}$
switches signs.

When $u>0$, the friction term leads to the local heating of the gas
if $1-\gamma uM^{2}/D>0$ and to cooling it otherwise. Because $f=c_{f}\rho u\left|u\right|$,
when $u\leq0$, we obtain that $1-\gamma uM^{2}/D>0$, but $f<0$.
Therefore, in the latter case, the friction cools the gas. Note that
this temperature decrease is local and is not that of a fluid particle.
For the latter, its rate of temperature change is $u\der T/\der\xi$,
the contribution to which by friction will remain positive when $u$
changes sign. 

\begin{figure}[H]
\noindent \begin{centering}
\includegraphics[height=4.43cm]{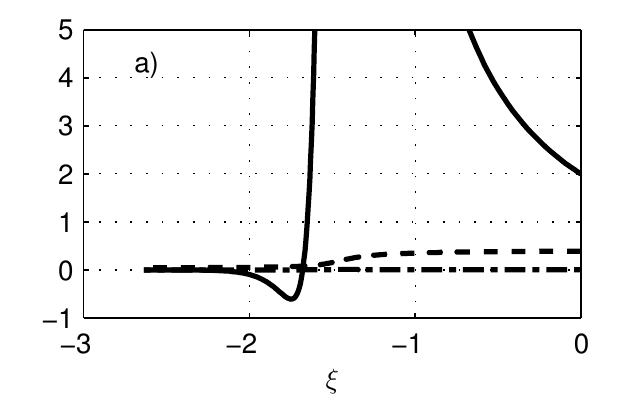}\includegraphics[clip,height=4.43cm]{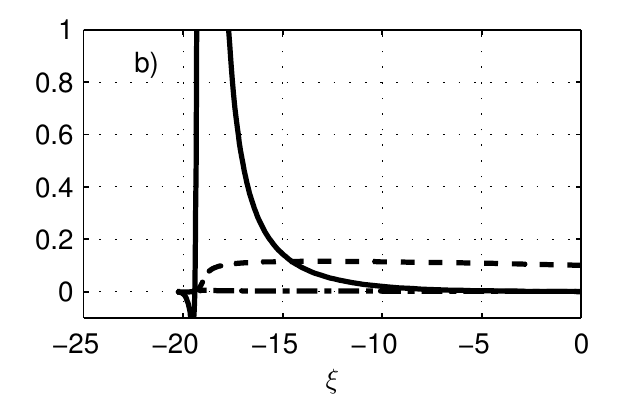}
\par\end{centering}

\caption{\label{fig:heating terms at large D} The profiles of various terms
contributing to the temperature evolution in (\ref{eq:T-eqn-detailed}),
$\mathcal{R}=\left(1-\gamma M^{2}\right)Q\omega$ (solid), $\mathcal{F}=D\left(1-\gamma uM^{2}/D\right)f/\left(\rho\phi\right)$
(dash) and $\mathcal{H}=\left(1-\gamma M^{2}\right)h/\left(\rho\phi\right)$
(dash-dot) at: a) $D/D_{\CJ}=0.95$ and b) $D/D_{\CJ}=0.6$.}
\end{figure}

Immediately at the shock, the jump conditions yield 
\[
\frac{\gamma u_{s}}{D}=\frac{\gamma u_{s}/c_{a}}{D/c_{a}}=\frac{2\gamma}{\gamma+1}\frac{M_{a}^{2}-1}{M_{a}^{2}}
\]
 and 
\[
M_{s}^{2}=\frac{2+\left(\gamma-1\right)M_{a}^{2}}{2\gamma M_{a}^{2}-\left(\gamma-1\right)}.
\]
A little bit of algebra shows that 
\begin{equation}
1-\frac{\gamma u_{s}}{D}M_{s}^{2}=\frac{4\gamma}{\left(\gamma+1\right)M_{a}^{2}\left(2\gamma M_{a}^{2}-\left(\gamma-1\right)\right)}\left[M_{a}^{4}-\frac{3\left(2-\gamma\right)}{4}M_{a}^{2}+1\right].
\end{equation}
Thus, the sign of the left-hand side is controlled by the expression
in the brackets, which is positive%
\footnote{With $t=M_{a}^{2}-1>0$, we find that $M_{a}^{4}-\frac{3\left(2-\gamma\right)}{4}M_{a}^{2}+1=t^{2}+\frac{14-3\gamma}{4}t+\frac{14-3\gamma}{4}$,
which is positive at $1<\gamma<14/3$.%
} for reasonable $\gamma$. Then, in the region immediately after the
shock, where $u>0$, the friction term always heats the gas. Again,
the important question here is to determine which of the two heating
processes, chemical or frictional, dominates. In order to answer this
question, we plot the terms in (\ref{eq:T-eqn-detailed}) that represent
various physics, at two different values of the detonation speed,
one close to $D_{\CJ}$ and one substantially below $D_{\CJ}$ as
shown in Fig. \ref{fig:heating terms at large D}. As expected, when
$D/D_{\CJ}=0.95$ the losses are small and the dominant role is played
by the reaction heating the gas and thus driving the detonation wave.
In contrast, at large velocity deficits, such as $D/D_{\CJ}=0.6$,
frictional heating is seen to dominate in the large region immediately
behind the shock. Comparison of the magnitudes of the friction terms
in (\ref{eq:T-eqn-detailed}) for these two cases shows that they
are substantially similar (approximately $0.4$ and $0.1$, the difference
following from an approximately two-fold decrease in velocity). The
reaction rate term, in contrast, decreases by orders of magnitude
in the vicinity of the shock, as a result of the exponential sensitivity
of the rate function to temperature. Thus, at low detonation velocities,
the dominant role of the frictional heating is a result of the reaction
becoming negligible. However, as the friction heats the gas, the reaction
``switches on'' at some distance from the shock. 

Thus, the propagation mechanism of low-velocity detonation involves
the shock-driven flow acceleration that leads to the frictional heating
of the gas that triggers the chemical energy release that, in turn,
drives the shock. The frictional heating in this mechanism is acting
as an intermediary facilitating the onset of chemical energy release
behind the shock, which itself is not strong enough to trigger the
reaction. We should note that the heat loss term in this early stage
of the process is negligible.

\begin{figure}[H]
\noindent \begin{centering}
\includegraphics[height=4.43cm]{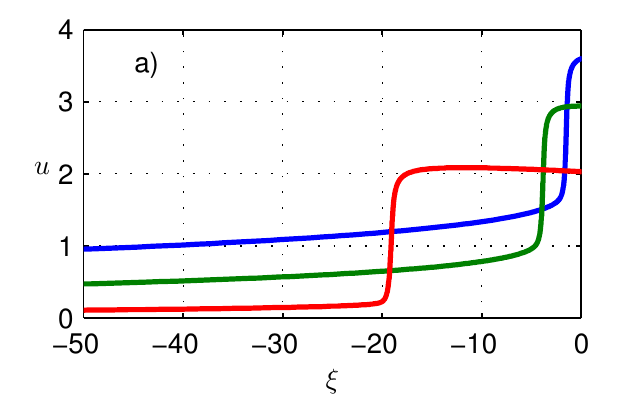}\includegraphics[clip,height=4.43cm]{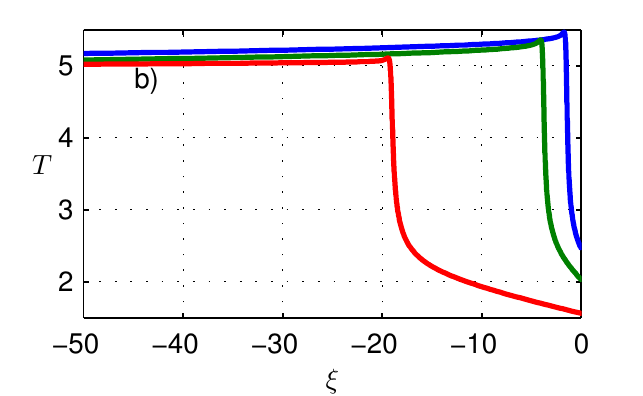}
\par\end{centering}

\noindent \begin{centering}
\includegraphics[height=4.43cm]{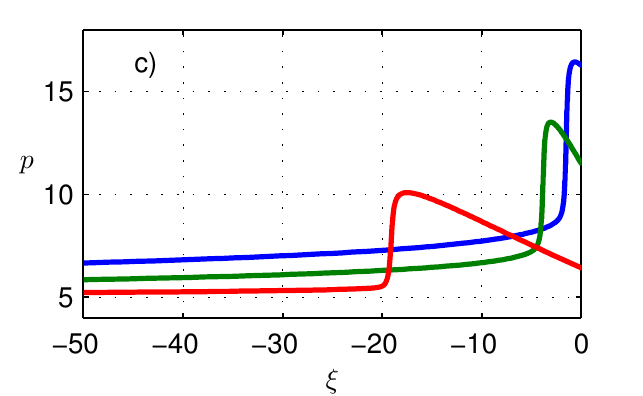}\includegraphics[clip,height=4.43cm]{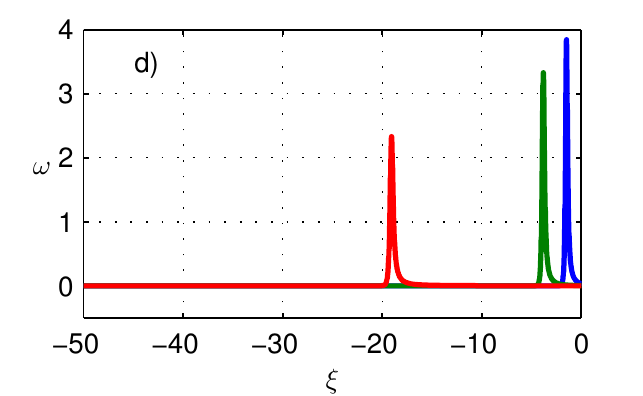}
\par\end{centering}

\noindent \begin{centering}
\includegraphics[height=4.43cm]{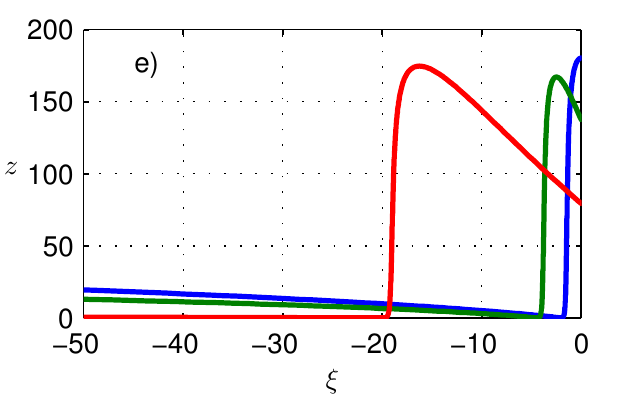}\includegraphics[clip,height=4.43cm]{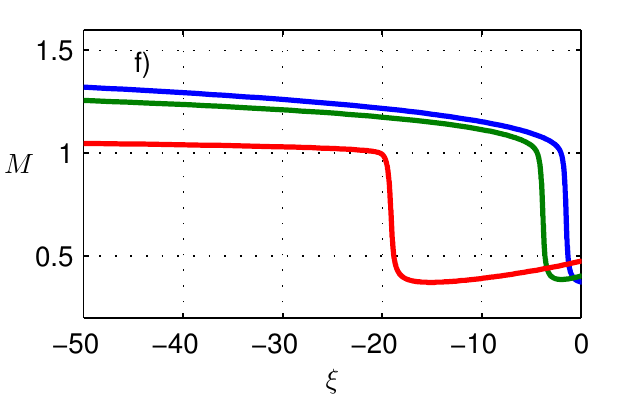}
\par\end{centering}

\caption{\label{fig:solution profiles at large D} The profiles of: a) velocity,
b) temperature, c) pressure, d) reaction rate, e) $z$ and f) Mach
number relative to the shock at: $D/D_{\CJ}=0.95$ (blue), $0.8$
(green) and $0.6$ (red). }
\end{figure}

\begin{figure}[H]
\noindent \begin{centering}
\includegraphics[width=10cm]{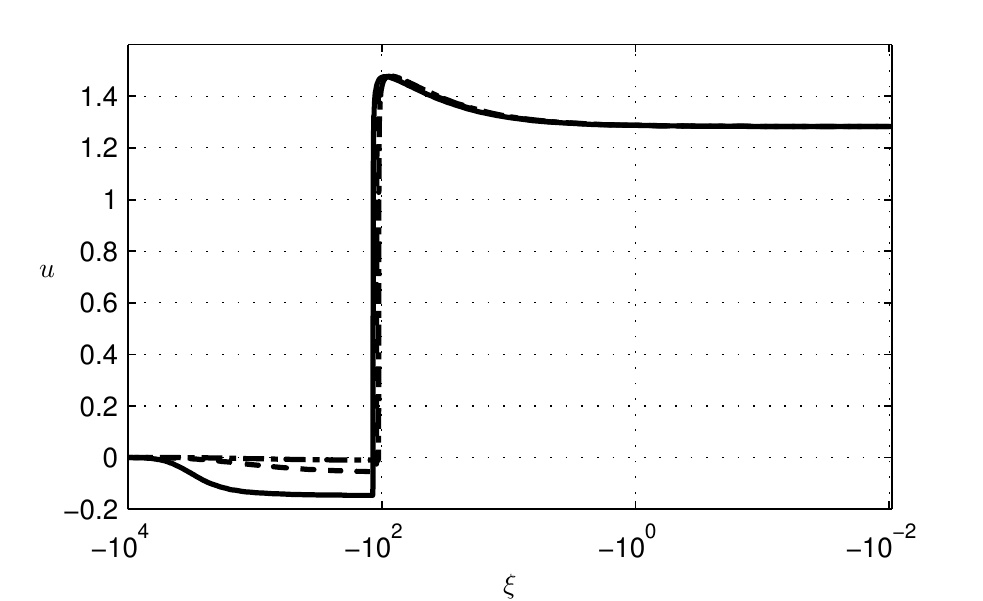}
\par\end{centering}

\caption{\label{fig:velocity-profile-log-scale} The velocity profiles at $D/D_{\CJ}=0.45$
and at three different values of $c_{f}$ in the set-valued region:
two near the edges of the region, $c_{f}=0.0019$ (solid), $c_{f}=0.0022$
(dash-dot) and one in the middle, $c_{f}=0.0021$ (dash).}
\end{figure}

In Fig. \ref{fig:solution profiles at large D}, we show the solution
profiles at three different levels of the velocity deficit. As the
deficit increases, the reaction is seen to start farther away from
the shock, without substantially changing the width of the heat release
zone (``fire'') with, however, a decreasing peak value of $\omega$.
There is a clear build-up of pressure and temperature between the
shock and the fire, which is due to frictional heating. The velocity
profile tends to a square-wave like structure with a nearly constant
value behind the shock that drops quickly in the fire zone to close
to zero. We should note that $u=0$ only at $\xi\to-\infty$ over
a long relaxation tail that extends from about $\xi\approx-20$. The
relaxation region is many orders of magnitude longer than the region
where most of the energy is released. We show an example in Fig. \ref{fig:velocity-profile-log-scale},
where the $\xi$ scale is logarithmic. It can be seen that the relaxation
of velocity from the negative values back to zero takes place over
a 100 times longer distance than the distance from the shock to the
fire. Although most of the figures below show only the region close
to the shock, in all of the results presented here, the calculations
are carried out to such long distances to verify that the velocity
has indeed reached a value close to zero. 

What is interesting and important in comparing the nature of the self-sustained
and set-valued solutions is the fact that the self-sustained solutions
do not require any boundary condition at infinity. In self-sustained
solutions, the velocity relaxes to $u=0$ automatically, once the
solution passes through the sonic point. It is therefore sufficient
to require that the sonic point exists in the flow. In contrast, $u\left(-\infty\right)=0$
has to be imposed as a boundary condition in solutions without the
sonic point. This observation highlights the special nature of self-sustained
solutions. The presence of a sonic point is a rather strong constraint
on a solution.

\begin{figure}[H]
\noindent \begin{centering}
\includegraphics[height=4.43cm]{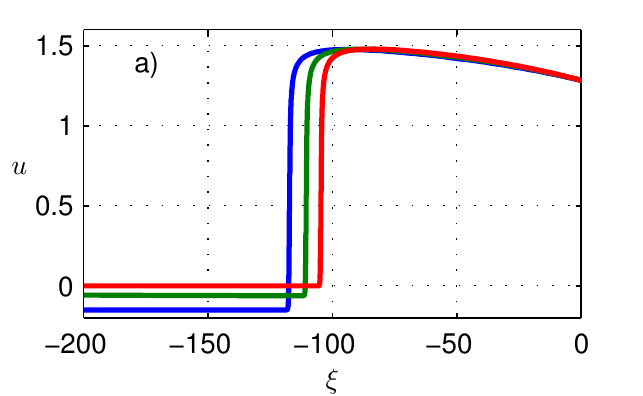}\includegraphics[clip,height=4.43cm]{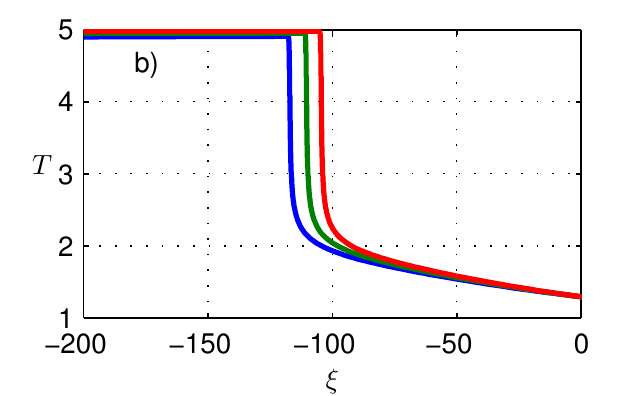}
\par\end{centering}

\noindent \begin{centering}
\includegraphics[height=4.43cm]{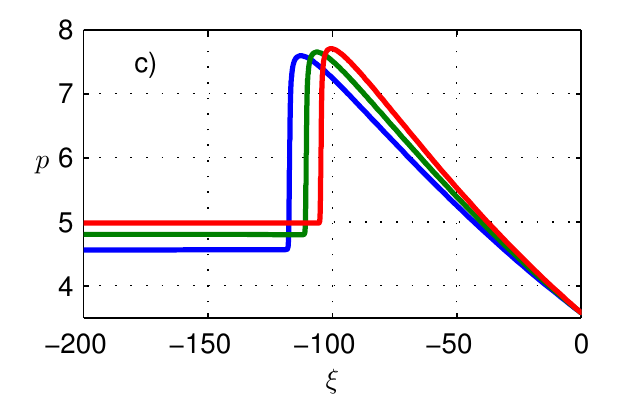}\includegraphics[clip,height=4.43cm]{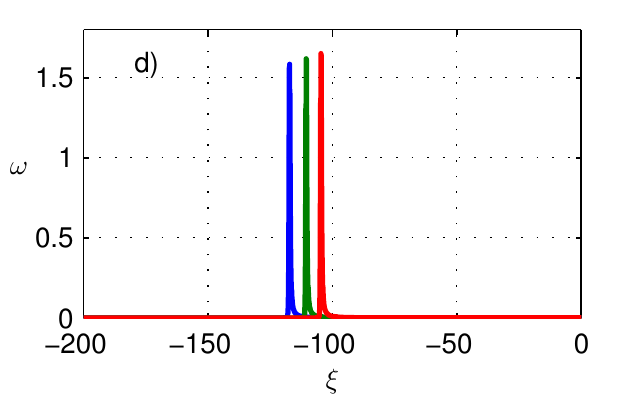}
\par\end{centering}

\noindent \begin{centering}
\includegraphics[height=4.43cm]{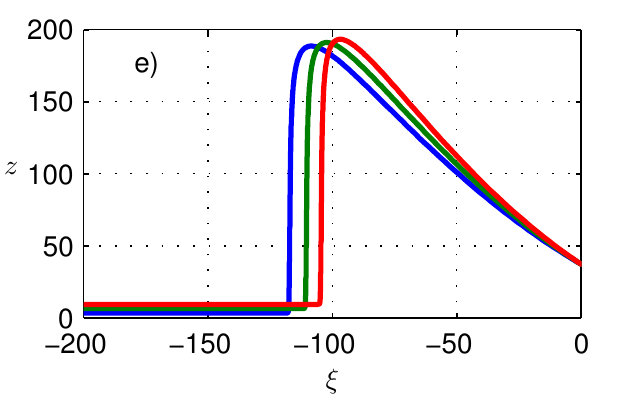}\includegraphics[clip,height=4.43cm]{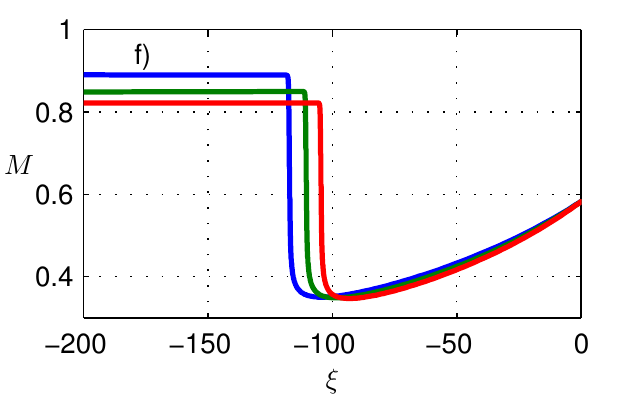}
\par\end{centering}

\caption{\label{fig:profiles-fixed-D-var-cf} The solution profiles at a fixed
$D/D_{\CJ}=0.45$ in the set-valued region: a) velocity, b) temperature,
c) pressure, d) reaction rate, e) $z$ and f) Mach number at three
different values of $c_{f}$: $c_{f}=0.0019$ (blue), $c_{f}=0.0021$
(green) and $c_{f}=0.0022$ (red). }
\end{figure}

\begin{figure}[H]
\noindent \begin{centering}
\includegraphics[height=5.43cm]{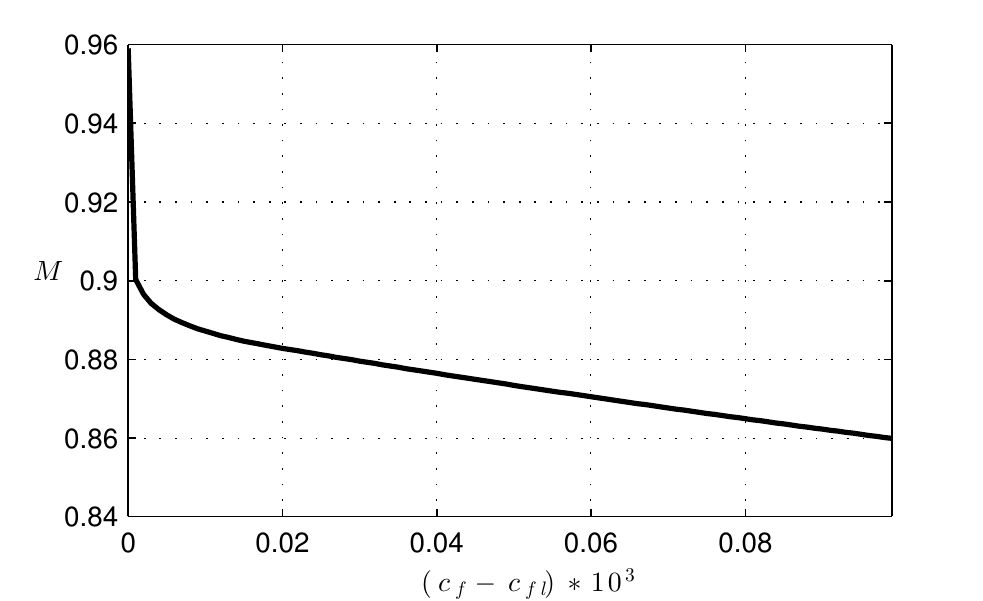}
\par\end{centering}

\caption{\label{fig:mach-at-infinity} The profile of $M_{b}=D/c\left(-\infty\right)$
as a function of $c_{f}$ at $D/D_{\CJ}=0.45$ as $c_{f}$ approaches
the left boundary of the set-valued region, $c_{fl}$. Within about
$10^{-8}$ of the boundary, the far-field Mach number remains below
$1$ although it does show a tendency towards $1$.}
\end{figure}

Next, we look at solutions without a sonic locus that exist at substantially
larger velocity deficits. In Fig. \ref{fig:profiles-fixed-D-var-cf},
we show the solution profiles when the velocity is $D=0.45D_{\CJ}$,
which is in the set-valued region. We take three values of $c_{f}$,
two near the boundaries of the set-valued region and one in between,
and plot the corresponding solution profiles. All of them look qualitatively
similar. The right boundary is characterized by the absence of the
negative phase of the velocity, which tends to zero asymptotically
as $\xi\to-\infty$. The approach to the left boundary is more complicated
and is shown in Fig. \ref{fig:mach-at-infinity}. In the set-valued
region, at values of $c_{f}$, which are away from the boundaries,
the local Mach number at infinity is $M_{b}=D/c\left(-\infty\right)<1$.
As we approach the left boundary of the region, $M_{b}$ remains substantially
below unity until we get very close to the left boundary of $c_{f}$,
which is denoted as $c_{fl}$. There is a sharp boundary layer at
$c_{fl}$ where $M_{b}$ starts to increase. We note that at $c_{f}<c_{fl}$,
the steady-state solution does not exist and the solutions of the
governing ODE system (for the original variables) blow up as a result
of $1-M^{2}$ becoming zero\textbf{. }This may explain the trend seen
just to the right of $c_{f}=c_{fl}$.

\begin{figure}[H]
\noindent \begin{centering}
\includegraphics[height=4.43cm]{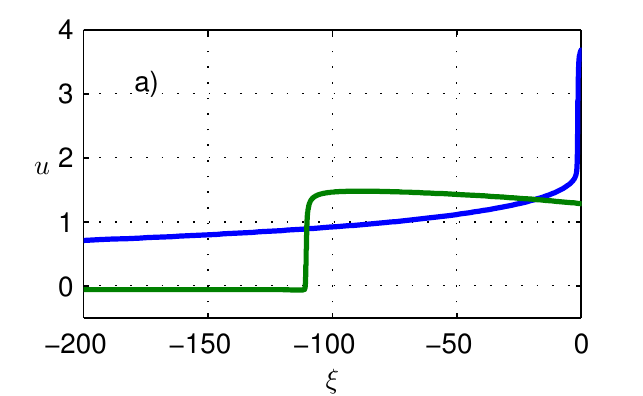}\includegraphics[clip,height=4.43cm]{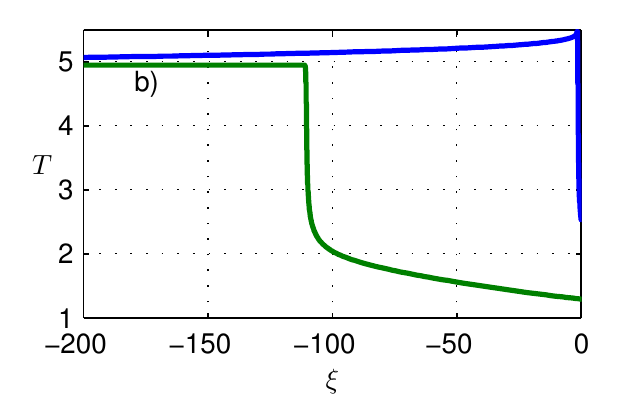}
\par\end{centering}

\noindent \begin{centering}
\includegraphics[height=4.43cm]{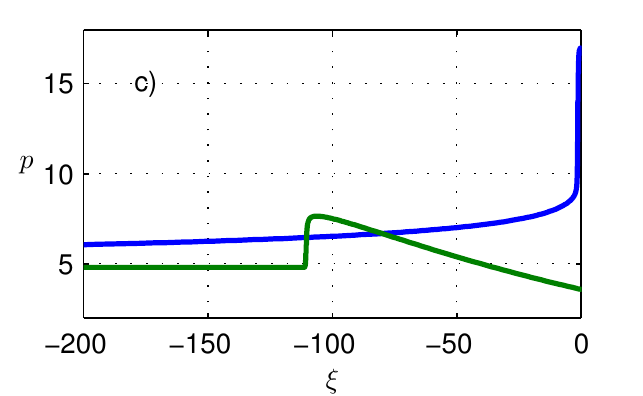}\includegraphics[clip,height=4.43cm]{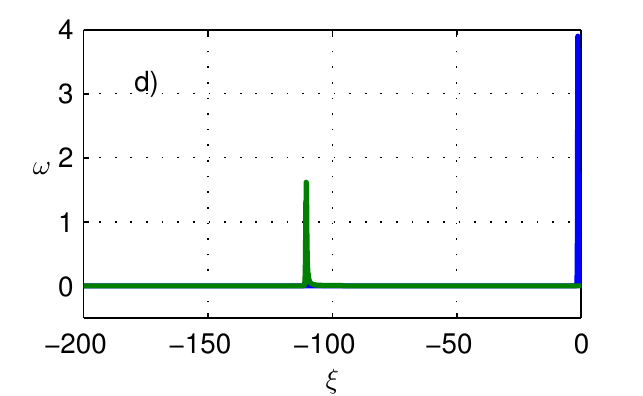}
\par\end{centering}

\noindent \begin{centering}
\includegraphics[height=4.43cm]{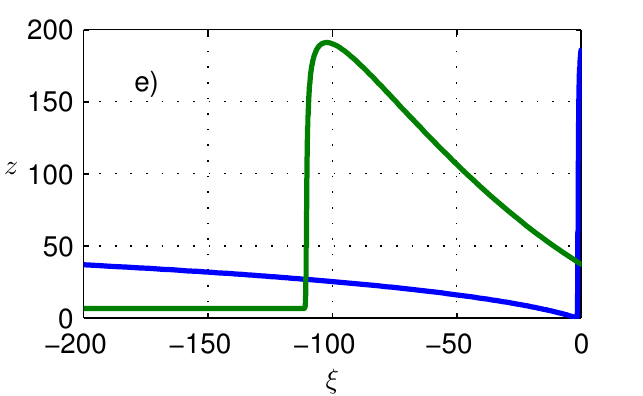}\includegraphics[clip,height=4.43cm]{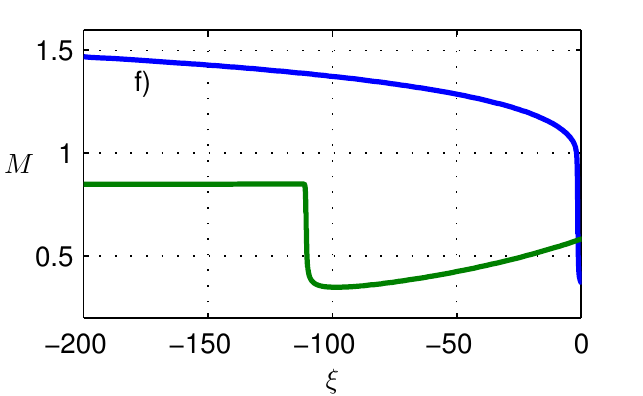}
\par\end{centering}

\caption{\label{fig:profiles-fixed-cf-var-D} The solution profiles at a fixed
$c_{f}=0.0021$ and: at $D/D_{\CJ}=0.97$ (blue) and at $D/D_{\CJ}=0.45$
(green). }
\end{figure}

In the next set of calculations, we look at the structure of the steady-state
solutions when $c_{f}$ is fixed and $D$ is allowed to take values
in the set-valued region as well as on the top branch of the $D$-$c_{f}$
dependence. In Fig. \ref{fig:profiles-fixed-cf-var-D}, we show such
profiles, which coexist for the same set of parameters describing
an explosive mixture and a porous medium. Comparing Figs. \ref{fig:solution profiles at large D},
\ref{fig:profiles-fixed-D-var-cf} and \ref{fig:profiles-fixed-cf-var-D},
we notice that the drop in detonation velocity at a given $c_{f}$
has a much more pronounced effect on the extent and the structure
of the solution profiles than the variations of $c_{f}$ at a given
velocity. 

The effect of the activation energy on the solutions is shown in Fig.
\ref{fig:D_cf curves for different E, gamma}a). The behaviour of
the top part of the dependence is as expected: as the activation energy
increases, the turning point tends to smaller values of $c_{f}$ and
larger values of $D$. With increasing $E$, the set-valued region
becomes significantly narrower, such that at a given $D$, the range
of $c_{f}$ for which the steady-state solution exists, is diminished.
However, note that for a given $c_{f}$ in the set-valued range, the
variations of $D$ are still significant, as the region becomes nearly
vertical. The latter is important in the sense that if one chooses
the porous medium to yield this particular value of $c_{f}$, then
the range of possible steady-state detonation velocities is still
large. 
\begin{figure}[H]
\noindent \begin{centering}
\includegraphics[clip,height=4.43cm]{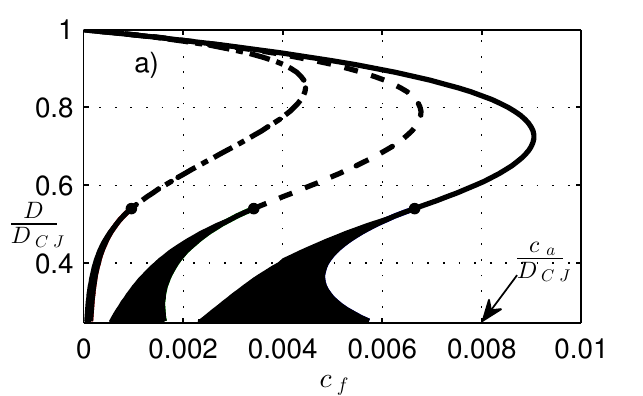}\includegraphics[clip,height=4.43cm]{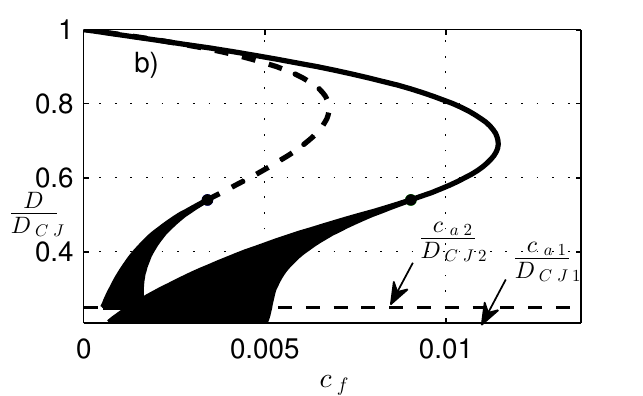}
\par\end{centering}

\caption{\label{fig:D_cf curves for different E, gamma} The effect of activation
energy and the ratio of specific heats: a) $\gamma=1.2$, $E=25$
(solid), $E=30$ (dash), and $E=40$ (dash-dot); b) $E=30$, $\gamma=1.2$
(solid) and $\gamma=1.3$ (dash).}
\end{figure}

The adiabatic index turns out to have a strong influence on the $D$-$c_{f}$
dependence as well as shown in Fig. \ref{fig:D_cf curves for different E, gamma}b).
Changing $\gamma$ from $1.2$ to $1.3$ shifts the dependence toward
smaller $c_{f}$ and the turning point toward larger $D$. In addition,
the set-valued part becomes narrower with increasing $\gamma$. The
bottom limit of the dependence changes as well due to the change in
the ambient sound speed.

\begin{figure}[H]
\noindent \begin{centering}
\includegraphics[clip,width=10cm]{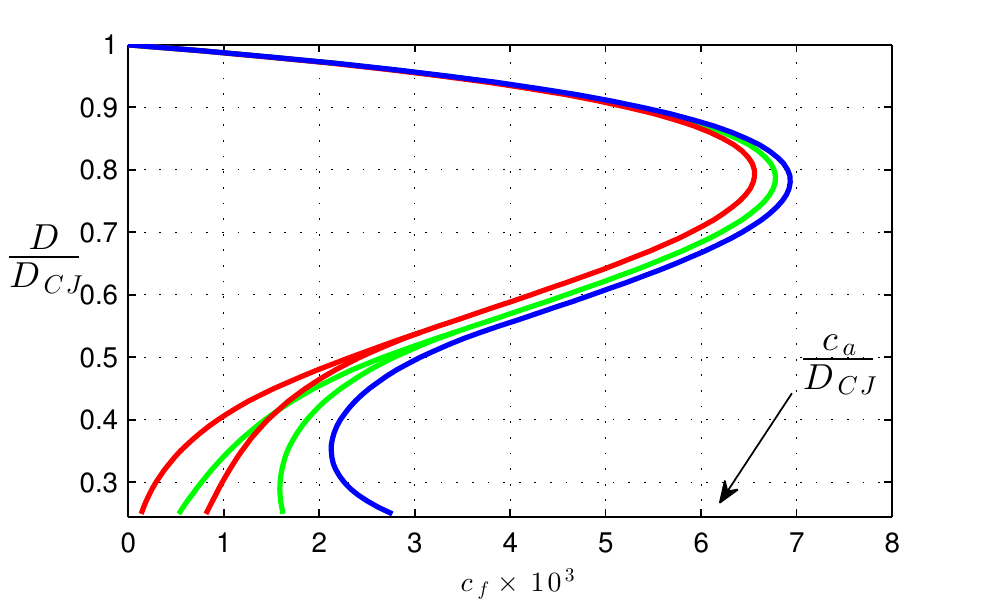}
\par\end{centering}

\caption{\label{fig:D_cf curves for different cf and ch ratio} The effect
of $\beta=c_{h}/c_{f}$: $\beta=0$ (blue), $\beta=0.4$ (green),
$\beta=1$ (red). The set-valued areas are not filled to avoid overlaps,
but it should be understood that the region between the red (correspondingly,
green) curves is set-valued. When $\beta=0$, the solution is a curve.
In all cases $E=30,\; Q=20,\;\gamma=1.2$.}
\end{figure}

The ratio $\beta=c_{h}/c_{f}$ is fixed at $0.4$ in most calculations
above. The self-sustained part of the solution is less affected by
the change of $\beta$ than is the set-valued part; this is shown
in Fig. \ref{fig:D_cf curves for different cf and ch ratio}. Note
that if the heat loss term is ignored, then no set valued solutions
exist and we obtain the familiar Z-shaped curve.

\section{\label{sec:Conclusions}Conclusions}

In this work, we revisited the classical problem of one-dimensional
gaseous detonation in the presence of losses of heat and momentum.
In particular, we analyzed the existence and structure of the steady-state
solutions of the one-dimensional reactive Euler equations for gaseous
detonation propagating in the interstitial space between packed spherical
particles. The friction force was assumed to be proportional to the
square of the flow velocity and the heat loss to the temperature difference
between the gas and the particles. The heat transfer was assumed to
be purely convective, such that it vanished in the absence of flow. 

Our main finding is that the steady-state solutions have a set-valued
character at sufficiently low speeds of detonation when the flow downstream
of the shock is entirely subsonic relative to the shock. The vanishing
velocity at the downstream infinity with no additional constraints
on the solution results in the existence of a continuous spectrum
of detonation velocities all corresponding to the steady-state solution
of the governing equations. For any given loss factors, such solutions
are found to exist provided that both the friction and heat losses
are considered and provided that the detonation velocity is sufficiently
low, i.e., the velocity deficit is large. This result, that the nonlinear
eigenvalue problem has in general both a discrete and a continuous
spectrum, is in contrast to previous findings for the problem, which
predicted only a finite number of discrete values of the detonation
speed at any given loss coefficient. 

Regarding the solutions with regions of negative velocity in the reaction
zone, we find that such regions cannot exist in self-sustained waves
(i.e. in the presence of the sonic point). They also do not exist
if one neglects heat losses, at least when only convective heat loss
is considered. Only in the presence of heat losses and only if the
solution contains no sonic point does the flow velocity take negative
values somewhere in the reaction zone. The latter situation corresponds
to the existence of the set-valued region in the $D-c_{f}$ plane.
The set-valued solutions arise because of the sonic constraint disappearing
from the solution and an additional degree of freedom appearing as
a result. This degree of freedom is the pressure or temperature at
infinity, which if imposed leads to the usual multiple but finite
number of solutions. This creates an interesting possibility that
the detonation speed can be controlled by the downstream pressure,
a situation akin to the overdriven case wherein a piston pushing the
products downstream controls the detonation velocity. However, the
difference in the present case is that the velocity is less than the
CJ velocity.

Another new result of this work is a formulation of governing equations
in terms of new dependent variables that yield an ODE system for steady-state
solutions without sonic singularity. The new formulation allows us
to eliminate completely the numerical difficulties associated with
the integration across the sonic singularity. Our approach can be
generalized and we believe will be useful in other, even more complex
problems of determining the detonation eigenvalue solutions. 

The existence of multiple steady-state solutions for the non-ideal
detonation raises an important question about their stability. Numerical
evidence points to a destabilizing role of losses, which is similar
to the role of curvature in expanding spherical/cylindrical detonations.
However, detailed theoretical study of stability of these solutions
is still required.

\bibliographystyle{plain}
\bibliography{/Users/aslankasimov/Dropbox/Biblioteka/akasimov-refs}

\end{document}